\documentclass[1p,10pt]{elsarticle}

\journal{\href{http://arxiv.org/}{arXiv.org}}

\usepackage{hyperref}
\usepackage{amssymb}
\usepackage{amsmath}
\usepackage{graphicx}
\usepackage[T1]{fontenc}

\begin{document}

\newcommand{\BR}{{\mathbb R}}
\newcommand{\BN}{{\mathbb N}}
\newcommand{\BZ}{{\mathbb Z}}
\newcommand{\BC}{{\mathbb C}}
\newcommand{\e}{{\bf e}}
\newcommand{\ajplus}{{a}_h(x_j)}
\newcommand{\ajminus}{{a}_h(x_j-h)}
\newcommand{\qj}{{\bf q}_h(x_j)}
\newcommand{\m}{{\bf m}}
\newcommand{\p}{{\bf p}}
\newcommand{\s}{{\bf s}}

\renewcommand{\Pr}{\mbox{Pr}}

\newcommand{\f}{{\bf f}}
\newcommand{\g}{{\bf g}}
\newcommand{\uh}{{\bf u}_h}

\newcommand{\cl}{C \kern -0.1em \ell}

\newcommand{\ov}{\overline}
\newcommand{\un}{\underline}
\newcommand{\proof}{\bf {Proof:} \rm}
\renewcommand{\qed}{$\blacksquare$}
\newtheorem{theorem}{Theorem}[section]
\newtheorem{remark}{Remark}[section]
\newtheorem{lemma}{Lemma}[section]
\newtheorem{problem}{Problem}[section]
\newtheorem{answer}{Answer}[section]
\newtheorem{proposition}{Proposition}[section]
\newtheorem{corollary}{Corollary}[section]
\newtheorem{definition}{Definition}[section]
\newtheorem{example}{Example}[section]

\begin{frontmatter}

\title{Hypercomplex Fock States for Discrete Electromagnetic Schr\"odinger Operators:
A Bayesian Probability Perspective}

\author{N.~Faustino\fnref{labelNelson}\corref{cor1}}
\ead{{nelson.faustino@ufabc.edu.br}}
\ead[url]{https://sites.google.com/site/nelsonfaustinopt/}
\cortext[cor1]{Corresponding author}
\fntext[labelNelson]{N. Faustino was formerly supported by the
fellowship \href{http://www.bv.fapesp.br/14523}{13/07590-8} of
FAPESP (S.P., Brazil).}
\address{
CMCC, Universidade Federal do ABC, 09210--580, Santo André, SP,
Brazil}


\begin{abstract}
We present and study a new class of Fock states underlying to
discrete electromagnetic Schr\"odinger operators from a multivector
calculus perspective. This naturally lead to hypercomplex versions
of Poisson-Charlier polynomials, Meixner polynomials, among other
ones. The foundations of this work are based on the exploitation of
the quantum probability formulation '\`a la Dirac' to the setting of
Bayesian probabilities, on which the Fock states arise as discrete
quasi-probability distributions carrying a set of {\it independent
and identically distributed} (i.i.d) random variables. By employing
Mellin-Barnes integrals in the complex plane we obtain counterparts
for the well-known multidimensional Poisson and hypergeometric
distributions, as well as quasi-probability distributions that may
take negative or complex values on the lattice $h\BZ^n$.

\end{abstract}

\end{frontmatter}

\tableofcontents


\section{Introduction}\label{Introduction}

  Discrete electromagnetic Schr\"odinger operators correspond to a
subclass of (doubly) Jacobi operators. They are ubiquitous in
several fields of mathematics, physics and beyond, as is witnessed
by the papers
\cite{FLtVinet93,GesztesyTeschl96,Simon98,VanDiejen05,RabinovichRoch09,
ChakJeugt10,CamposKravchenko11,StoilovaJeugt11,MikiTsuVinetZhed12,AptekarevDerevVanAssche14}
and monograph \cite{Teschl00}. Here, the factorization method is the
cornerstone in the study of the quasi-exact solvability of such kind
of operators since it avoids non-perturbative arguments that appear
under the discretization of its {\it continuum} counterpart,
 the quantum harmonic oscillator
$-\frac{1}{2m}\Delta+V(x)$ with mass $m$ and potential $V(x)$
(cf.~\cite{FLtVinet93,SpVinetZhedanov93}). In case of
crystallographic root systems are involved, the discrete
electromagnetic Schr\"odinger operators may be described as discrete
(pseudo) Laplacians (cf.~\cite[Section 6]{VanDiejen05}), whose
origin goes back to the works of Macdonald
\cite{Macdonald96,Macdonald00}. As it was shown in Ruijsenaars's
seminal work \cite{Ruijsenaars02}, Macdonald's theory may be
obtained as a special case of integrable lattice models of
Calogero-Moser type that exhibit factorized scattering. For further
details, we refer to \cite{Ruijsenaars95}.

The main objective of this paper is to show the feasibility of
special functions of hypercomplex variable, with values on the
Clifford algebra of signature $(0,n)$, as Fock states of a certain
multidimensional Schr\"odinger operator $L_h$ acting on the lattice
$$\displaystyle h\BZ^n=\left\{ (x_1,x_2,\ldots,x_n) \in
\BR^n~:~\frac{x_j}{h}\in \BZ~,~\mbox{for
all}~j=1,2,\ldots,n\right\},$$ with mesh width $h>0$.

In the series of papers \cite{FR11,Faustino13,FaustinoMonomiality14}
the author developed a framework to compute, in a direct manner,
quasi-monomials of discrete hypercomplex variable from the knowledge
of a underlying set of Lie-algebraic symmetries. The methods and
techniques employed are closely related with Wigner's quantal
systems and go far beyond the symmetries of the Weyl-Heisenberg
algebra, mentioned in many textbooks as the underlying symmetries
encoded by Hermite polynomials/functions (cf.~\cite{CFK11}).

In this paper we center our analysis on questions regarding the
quasi-exact solvability associated to a discretization $L_h$ of a
Sturm-Liouville type operator. This essentially corresponds to the
problem formulation:
\begin{problem}\label{ExactSolvabilityProblem}
 Given a pair of Clifford-vector-valued operators
$(A_h^+,A_h^-)$ satisfying $$L_h=\frac{1}{2}\left(A_h^+A_h^- + A_h^-
A_h^+\right),$$ can we recover the discrete electric and magnetic
potentials of $L_h$, $\Phi_h(x)$ and ${\bf a}_h(x)$ respectively,
from the knowledge of its $k-$Fock states $\psi_k(x;h)$ ($k\in
\BN_0$)?
\end{problem}

 Here, the construction of the pair $(A_h^+,A_h^-)$
  was inspired on Spiridonov-Vinet-Zhedanov approach \cite{SpVinetZhedanov93} and
roughly follows the same order of ideas used on Odake-Sasaki's
papers
\cite{OdakeSasaki05,OdakeSasaki09,OdakeSasaki09b,OdakeSasaki10} to
generate one-dimensional 'discrete' quantum systems through the {\it
Supersymmetric Quantum Mechanics} (SUSY QM)\footnote{The
fundamentals of SUSY QM can be traced back to the seminal work of
Cooper-Khare-Sukhatme \cite{CKS1995}, where the interest lies
essentially in the solution of Pauli and Dirac equations
(cf.~\cite[Section 10. \& Section 11.]{CKS1995}).} framework.

We are not concerned here with a SUSY QM extension/generalization to
hypercomplex variables in the way that the $k-$Fock states are
eigenfunctions of one of the Hamiltonians, $A_h^-A_h^+$ and
$A_h^+A_h^-$ respectively, neither with an exploitation of the
commutation method (cf.~\cite[Chapter 11]{Teschl00}). On the context
of this paper, the $k-$Fock states $\psi_k(x;h)$ shall be understood
as basis functions with membership in a certain linear subspace
$\mathcal{F}_h$ of the Hilbert module
$\ell_2(h\BZ^n;\cl_{0,n})=\ell_2(h\BZ^n)\otimes \cl_{0,n}$,
generated from $(A_h^+,A_h^-)$ -- the so-called Fock space
$\mathcal{F}_h$, to be defined later on this paper.

 Of particular
importance for the development of this approach will be the
connection with Bayesian probabilities that results from the
observation that, for a given {\it ground state} $\psi_0(x;h)$
satisfying $\langle \psi_0,\psi_0 \rangle_h=1$, the quantity
\begin{eqnarray}
\label{quasiProbability}\Pr\left(\sum_{j=1}^n \e_j X_j=x
\right)=h^n\psi_0(x;h)^\dag \psi_0(x;h)
\end{eqnarray}
may be regarded as a discrete quasi-probability law on $h\BZ^n$,
carrying a set of {\it independent and identically distributed}
(i.i.d.) random variables $X_1,X_2,\ldots,X_n$.

This quasi-probability formulation is reminiscent of a similar
probability formulation, considered in the context of transition
probabilities (cf.~\cite{CavesFuchsSchack02,Mouayn14}). In that
scope, the Bayesian scheme is achieved to determine the expectation
values of quantum observables, which are essentially the Landau
levels attached to the discrete electromagnetic Schr\"odinger
operator (\ref{JacobiOperator}) when one considers the minimization
problem
\begin{eqnarray*}
\label{minProblem} \displaystyle \psi=
\mbox{argmin}_{\widetilde{\psi}} \dfrac{\langle \widetilde{\psi},
L_h\widetilde{\psi} \rangle_h} {\langle
\widetilde{\psi},\widetilde{\psi} \rangle_h }
\end{eqnarray*}
to seek the quantum state $\psi$ with 'best energy concentration' in
$h\BZ^n$.

Accordingly to the general theory, in case that $L_h$ is real-valued
and symmetric -- the so-called Hermitian condition -- is sufficient
to guarantee that $L_h$ is quasi-exactly solvable (cf.
\cite[Proposition 1.4]{Simon98}). That's indeed the case of the
characterization provided through the formulation of {\bf Problem
\ref{ExactSolvabilityProblem}} (see also
\ref{ExactSolvabilityLemma}). Surprisingly enough, Bender and its
collaborators have been stressed in a series of papers (see
\cite{Bender05,BenderEtAll06,BenderHook08} and the references given
there) that such condition is not necessary\footnote{See also the
examples treated in \cite[Subsection 12.1.]{CKS1995} and in
\cite[Section 6]{AptekarevDerevVanAssche14}.} and may be replaced
with a most general one, involving a space-time reflection symmetry
(shorty, a $\mathcal{PT}$ symmetry) constraint. Thus, it may happen
that the right-hand side of (\ref{quasiProbability}) may also take
complex values (cf.~\cite{BenderHookMeiWang10}).

To be in accordance with Dirac's insight \cite{Dirac42} on quantum
probabilities, we will consider throughout this paper the
$\dag-$operation provided by (\ref{conjugation}), also for bound
states that take values in the complexified Clifford algebra $\BC
\otimes \cl_{0,n}$. We turn next to the content and the organization
of the subsequent sections:

\begin{itemize}
\item In {\bf Section \ref{TheSetting}}
 we will introduce the basic setting that will be used throughout the
 paper, namely the multivector calculus embody in the Clifford algebra
 $\cl_{0,n}$ in the spirit of Sturm-Liouville theory. We will also introduce some basic
features in the context of Fock spaces (cf.~\cite{Fock32}) to
describe the Fock states of the discrete electromagnetic
Schr\"odinger operator $L_h$ on $h\BZ^n$.
\item In {\bf Section \ref{BoundSection}} we will take into account the
factorization of $L_h$ and the {\it vacuum} vector $\psi_0(x;h)$ of
$A_h^+$ to display a correspondence between the Fock states of the
form $\psi_k(x;h)=\left(A_h^-\right)^k\psi_0(x;h)$ and the
quasi-monomials, encoded by the pair $(D_h^+,M_h)$, where $D_h^+$
stands the finite difference Dirac operator of forward type. That
corresponds to Lemma \ref{quasiMonomialLemma}. Moreover, with
Proposition \ref{AnswerMagnetic} and Proposition
\ref{AnswerMagneticM1} we hereby provide an answer to {\bf Problem
\ref{ExactSolvabilityProblem}}.
\item In {\bf Section \ref{BayesianProbabilitySection}} we will make use of the
Bayesian probability framework beyond Dirac's insight \cite{Dirac42}
to compute some examples involving the well-known Poisson and
hypergeometric distributions, likewise quasi-probability
distributions involving the generalized Mittag-Leffler/Wright
functions.
\item In {\bf Section \ref{RemarksQuasiP}} we conclude
with a more detailed discussion of Bayesian probabilities with {\it
imaginary} bias, towards the regularization of the Mittag-Leffler
distribution.
\item In {\bf Section \ref{ConclusionsSection}} we will outlook the main
contributions obtained and will raise some problems/questions to be
investigated afterwards.
\end{itemize}

\section{The Setting}\label{TheSetting}

We start this section by collecting some basic facts about Clifford
algebras that will be used on the sequel. We refer \cite[Chapter
1]{GilbertMurray91} for further details.

 Recall that $\cl_{0,n}$ is the algebra generated by the
set of vectors $\e_1,\e_2,\ldots,\e_n$ that satisfy, for each
$j,k=1,2,\ldots,n$, the set of anti-commuting relations
\begin{eqnarray}
\label{CliffordGenerators} \e_j\e_k+\e_k\e_j=-2\delta_{jk}.
\end{eqnarray}

The Clifford algebra $\cl_{0,n}$ is an associative algebra with
identity $1$ and dimension $2^n$, that contains $\BR$ and $\BR^n$ as
subspaces. This in particular means that for two given $n-$tuples
$(x_1,x_2,\ldots,x_n)$ and $(y_1,y_2,\ldots,y_n)$ of $\BR^n$,
represented on $\cl_{0,n}$ through the linear combinations
\begin{center}
$\displaystyle x=\sum_{j=1}^nx_j\e_j$ and
$\displaystyle y=\sum_{j=1}^ny_j\e_j$,
\end{center} respectively, the anti-commutator
$\displaystyle xy+yx=-2\sum_{j=1}^n x_jy_j$ is scalar-valued.

We will use throughout this paper the notations $\displaystyle
\mathcal{B}(x,y)=-\frac{1}{2}(xy+yx)$ to denote the bilinear form of
$\BR^n$ and $x\pm h\e_j$ to denote the underlying forward/backward
shifts $(x_1,x_2,\ldots,x_j\pm h,\ldots,x_n)$ on $h\BZ^n$. Generally
speaking, on $\cl_{0,n}$ one may consider for a subset
$J=\{j_1,j_2,\ldots,j_r\}$ of $\{ 1,2,\ldots,n\}$, with $1\leq
j_1<j_2<\ldots<j_r\leq n$, $r$-multivector bases of the form
$\e_J=\e_{j_1}\e_{j_2}\ldots \e_{j_r}$, and moreover,
Clifford-vector-valued functions $\f(x)$ as linear combinations of
the above form
 \begin{eqnarray*}
\f(x)=\sum_{r=0}^n\sum_{|J|=r} f_J(x) ~\e_J, & \mbox{with} &
f_{J}(x)~\mbox{scalar-valued}.
\end{eqnarray*}
Hereby $|J|$ denotes the cardinality of $J$. The $\dag-${\it
conjugation} operation $\f(x)\mapsto\f(x)^\dag$, defined as
 \begin{eqnarray}
 \label{conjugation}
\f(x)^\dag=\sum_{r=0}^n\sum_{|J|=r} f_J(x) ~\e_J^\dag, & \mbox{with}
& \e_J^\dag=(-1)^{r}\e_{j_r} \ldots \e_{j_2}\e_{j_1}
\end{eqnarray}
is an automorphism of $\cl_{0,n}$ satisfying, for each $\f(x)$ and
$\g(x)$, the conjugation properties
\begin{eqnarray}
\label{dagConjugation}
\left(\f(x)^\dag\right)^\dag=\f(x)&\mbox{and}&
\left(\f(x)\g(x)\right)^\dag=\g(x)^\dag \f(x)^\dag.
\end{eqnarray}

 The conjugation
properties on $\cl_{0,n}$ are two-fold since they correspond to a
generalization of the standard conjugation in the field of complex
numbers and to the multivector extension of the Hermitian
conjugation operation in the scope of matrix theory. In particular,
it follows from the property $\e_j^\dag=-\e_j$ and from the
anti-commutator relations (\ref{CliffordGenerators}) that the
quantities $\f(x)^\dag \f(x)$ and $\f(x)\f(x)^\dag $ are
scalar-valued and coincide. Being $\displaystyle \f(x)=\sum_{j=1}^n
f_j(x)\e_j$ a Clifford vector representation of the vector-field
$(f_1(x),f_2(x),\ldots,f_n(x))$ of $\BR^n$, one readily has
$$\f(x)^\dag \f(x)=\f(x)\f(x)^\dag=\sum_{j=0}^n
f_j(x)^2,$$ which is nothing else than the square of the Euclidean
norm on $\BR^n$.

Moreover, if we use the $\dag-$conjugation operator also for
functions $\f(x)$ with values in the complexified Clifford algebra
$\BC\otimes \cl_{0,n}$, the resulting quantity $\f(x)^\dag \f(x)$
must be interpreted, on a wide generality, as a quasi-probability in
Dirac's sense, i.e. a {\it probability density function} that
encompasses negative values (cf.~\cite[p.~4]{Dirac42}).

From the above characterization, one can thereafter define the set
of Clifford-vector-valued functions $\f(x)$ on $h\BZ^n$ with
membership on the Hilbert module
$\ell_2(h\BZ^n;\cl_{0,n})=\ell_2(h\BZ^n)\otimes \cl_{0,n}$, as the
linear space endowed by the sesquilinear form
\begin{eqnarray*}
\label{l2Inner} \langle \f,\g \rangle_h=\sum_{x\in h\BZ^n}h^n~
\f(x)^\dag\g(x).
\end{eqnarray*}

The class of discrete electromagnetic Schr\"odinger operators on
$h\BZ^n$ that we will consider throughout this paper are defined
{\it viz}
\begin{eqnarray}
\label{JacobiOperator} L_h \f(x)= \frac{1}{2\mu}\sum_{j=1}^n
\left(\frac{2}{qh}\f(x)-\ajplus\f(x+h\e_j)-\ajminus\f(x-h\e_j)~\right)
+q~\Phi_h(x)\f(x).
\end{eqnarray}

Hereby $\Phi_h(x)$ (scalar-valued function) denotes
 the discrete analogue of the electric potential
whereas $\displaystyle {\bf a}_h(x)=\sum_{j=1}^n \e_j\ajplus$
(vector-valued function) denotes the discrete analogue of the
magnetic potential. The parameters $\mu$ and $q$ denote the mass and
 the electric charge of the electron, respectively.
In case of $\Phi_h(x)$ and ${\bf a}_h(x)$ satisfy the set of
constraints
\begin{eqnarray*}
\displaystyle q\Phi_h(x)+\frac{1}{2\mu}\sum_{j=1}^n
\left(\frac{2}{qh}-\ajplus-\ajminus\right)=V\left(\frac{x}{h}\right)
+O\left(h^3\right) \\
  {\bf a}_h(x)=\sum_{j=1}^n\e_j~
  w\left(\frac{1}{q}~\frac{x_j}{h}\right)\left(1+O\left(h\right)\right),
\end{eqnarray*}
where $w\left(\frac{x_j}{qh}\right)$ stands the leading term of
$a_h(x_j)$, one gets (cf.~\cite[p.~64]{SpVinetZhedanov93})
\begin{eqnarray}
\label{Hamiltonian} L_h \f(x)=
 -\frac{h^2}{2\mu}\sum_{j=1}^n
 \frac{\partial}{\partial x_j}\left(w\left(\dfrac{x_j}{qh}\right)\frac{\partial \f}{\partial x_j}(x)\right)
 +V\left(\frac{x}{h}\right)\f(x)+O\left(h^3\right).
\end{eqnarray}

The above asymptotic model may be seen as a discrete counterpart of
the Sturm-Liouville operator. Here one notice that the exact
solvability of the right-hand side of (\ref{Hamiltonian}) was
studied in detail on \cite[Section 2]{CKS1995}. For the case of
$\ajplus \sim \dfrac{1}{qh}$ (~i.e.
$w\left(\frac{1}{q}~\frac{x_j}{h}\right)=\dfrac{1}{qh}$), $L_h$ is
asymptotically equivalent to the discrete harmonic oscillator
$-\frac{1}{2m}\Delta_h+q\Phi_h(x)$ with mass $m\sim \dfrac{\mu
q}{h}$, whose kinetic term is written in terms of the star Laplacian
(cf.~\cite[p.~423]{VanDiejen05},~\cite[p.~1967]{FR11} \&
\cite[p.~4]{RabinovichRoch09})
$$\Delta_h \f(x)=\sum_{j=1}^n \dfrac{\f(x+h\e_j)+\f(x-h\e_j)-2\f(x)}{h^2}.$$

Other interesting examples may arise if
$w\left(\frac{1}{q}~\frac{x_j}{h}\right)$ has polynomial behavior,
exponential behaviour or even if one takes a e.g. quantum
deformation of the constant polynomial
$w\left(\frac{1}{q}~\frac{x_j}{h}\right)=\dfrac{\mu}{qh}$
(cf.~\cite[Section 2.]{VanDiejenEmsiz15AnnHenriP}). For the
particular choice $a_h(x_j)= \dfrac{1}{q}~\left(\frac{1}{h}+\mu
\frac{x_j}{h}+\dfrac{\mu}{2}\right)$ it readily follows from the set
of identities
\begin{eqnarray*}
\begin{array}{ccc}
\f(x+h\e_j)&=&\f(x)+h\left(\dfrac{\f(x+h\e_j)-\f(x)}{h}\right) \\
\f(x-h\e_j)&=&\f(x)-h\left(\dfrac{\f(x)-\f(x-h\e_j)}{h}\right)
\end{array} & (j=1,2,\ldots,n)
\end{eqnarray*}
that the asymptotic expansion of the discrete electromagnetic
Schr\"odinger operator (\ref{JacobiOperator}) reduces to
\begin{eqnarray*}
L_h \f(x)= -\dfrac{1}{2\mu
q}(E_h^+\f(x)-E_h^-\f(x))+V\left(\frac{x}{h}\right)\f(x),
\end{eqnarray*}
with $\displaystyle
V\left(\frac{x}{h}\right)=-\sum_{j=1}^n\frac{x_j}{h} +q\Phi_h(x)$.
 Hereby
$$E_h^\pm\f(x)
=\sum_{j=1}^n\left(\frac{1}{\mu}+x_j\pm\frac{h}{2}\right)\dfrac{\f(x\pm
h\e_j)-\f(x)}{\pm h}$$ corresponds to the forward/backward
counterpart\footnote{The formulation of $E_h^\pm$ introduced in
\cite{Faustino13} is seemingly close to the formulation of number
operator in quantum mechanics and it goes beyond the standard
discretizations of the Euler operator $E$.} of the so-called Euler
operator $\displaystyle E=\sum_{j=1}^n x_j
\partial_{x_j}$, carrying the polynomial $w(x_j)=1+\mu x_j$
(cf.~\cite[p.~3]{Faustino13}). That gives us an alternative way to
discretize the quantum harmonic oscillator
$-\frac{1}{2m}\Delta+V\left(\frac{x}{h}\right)$ with mass term $m
\sim \dfrac{\mu q}{h}$.

\bigskip

 Next, we will consider a faithful adaptation of the Fock
space formalism \cite{Fock32}, already considered in \cite{CFK11},
to discrete hypercomplex variables. We introduce the Fock space
structure over $h\BZ^n$ as a linear subspace $\mathcal{F}_h$ of
$\ell_2(h\BZ^n;\cl_{0,n})$ encoded by the pair $(A_h^+,A_h^-)$ of
Clifford-vector-valued operators. To be more precise, we say that
$\mathcal{F}_h$ defines a Fock space over $h\BZ^n$ if the following
conditions are satisfied:
\begin{enumerate}
 \item {\bf Duality condition:} For two given lattice functions $\f(x)$ and $\g(x)$
 with membership in $\mathcal{F}_h$, the pair of Clifford-vector-valued operators
$(A_h^+,A_h^-)$ satisfy
 $$\langle
A_h^+ \f,\g \rangle_h=
 \langle \f,A_h^-\g \rangle_h.$$
\item {\bf Vacuum vector condition:} There exists a lattice
function $\psi_0(x;h)$ with membership in $\mathcal{F}_h$ such that
$$A_h^+\psi_0(x;h)=0.$$
 \item {\bf Energy
condition:} The {\it vacuum} vector $\psi_0$ satisfies
$$\langle \psi_0,\psi_0 \rangle_h=1.$$
\end{enumerate}

From direct application of the quantum field lemma
(cf.~\cite{Fock32}) the resulting Fock space $\mathcal{F}_h$ is thus
generated by the $k-$Fock states
\begin{eqnarray}
\label{PhikXh} \psi_k(x;h)=(A_h^-)^k\psi_0(x;h).
\end{eqnarray}

It readily follows from the $\dag-$conjugation property
(\ref{dagConjugation}) that the left representation $\Lambda(\s):
\f(x) \mapsto \s \f(x)$ is an isometry on $\ell_2(h\BZ^n;\cl_{0,n})$
whenever $\s\s^\dag=\s^\dag \s=1$, i.e.
\begin{eqnarray}
\label{unitaryRepConstraint} \langle \s
\f(x),\s\g(x)\rangle_h=\langle \f(x),\g(x)\rangle_h.
\end{eqnarray}

Regarding the above isometry property one may consider the Lie
groups $O(n)$ and $SO(n)$. Here $O(n)$ is the group of linear
transformations of $\BR^n$ which leave invariant the bilinear form
$\displaystyle \mathcal{B}(x,y)=-\frac{1}{2}(xy+yx)$ and $SO(n)$
(the so-called {\it special orthogonal group}) is the group of
linear transformations with determinant $1$. These groups have
natural transitive actions on the $(n-1)-$sphere
$$S^{n-1}=\left\{x=\sum_{j=1}^n x_j\e_j\in \cl_{0,n}~:~x^\dag x=x
x^\dag=1 \right\}.$$

Namely, through the action of $SO(n)$ we can rewrite every
$x\in\BR^n$ as $x=\rho\s$, with $\rho=\dfrac{x}{|x|}$ and $\s \in
S^{n-1}$. Using the fact that the group stabilizer of the Clifford
vector $\e_n\in \cl_{0,n}$ is isomorphic to $SO(n-1)$, the points of
$\s$ of $S^{n-1}$ can be identified with the homogeneous space
$SO(n)/SO(n-1)$ through the isomorphism property $SO(n)/SO(n-1)\cong
S^{n-1}$. In terms of the main involution operation $\s \mapsto
\s'$, defined in $\cl_{0,n}$ as
\begin{eqnarray*}
\label{involution} \s'=\sum_{r=0}^n\sum_{|J|=r}s_J \e_J'&
\mbox{with}& \e_J'=(-1)^r\e_{j_1}\e_{j_2}\ldots {\e_{j_r}},
\end{eqnarray*}
we find that the $\mbox{Pin}$ group $$\displaystyle
\mbox{Pin}(n)=\left\{ \s=\prod_{p=1}^q \s_p ~:~\s_1,\s_2,\ldots,
\s_q\in S^{n-1}, ~q\in\BN\right\}$$ and the $\mbox{Spin}$ group
$$\displaystyle \mbox{Spin}(n)=\left\{ \s=\prod_{p=0}^{2q} \s_p
~:~\s_1,\s_2,\ldots, \s_{2q}\in S^{n-1}, ~q\in\BN\right\}$$ may be
regarded as the underlying double-covering sheets for the groups
$O(n)$ and $SO(n)$, respectively, endowed by homomorphism action
$\chi(\s):\f(x)\mapsto \s \f(x) (\s')^{-1}$ (cf. \cite[Chapter
3]{GilbertMurray91}).

 Since $\mbox{Spin}(n)$
is a subgroup of $\mbox{Pin}(n)$, it remains natural to look
throughout for {\it vacuum} vectors $\psi_0(x;h)$ of the form
$\psi_0(x;h)=\phi(x;h)\s$, where $\phi(x;h)$ is scalar-valued and
$\s \in \mbox{Pin}(n)$. From now on we will always use the bold
notation $\s$ when we are referring to an element of $\mbox{Pin}(n)$
or $\mbox{Spin}(n)$.

\section{Main Results}\label{BoundSection}

\subsection{Factorization Approach}

We now turn to the factorization question posed in {\bf Problem
\ref{ExactSolvabilityProblem}}. To do so, we consider the set of
operators, $A_h^+$ and $A_h^-$, defined {\it viz}
\begin{eqnarray} \label{Ahpmqh}  \begin{array}{lll}
\displaystyle A_h^+=\sum_{j=1}^n \e_j A_h^{+j} &
\mbox{with}&A_h^{+j}=\sqrt{\dfrac{qh}{4\mu}}\left(\ajplus
T_h^{+j}-\dfrac{2}{qh}I\right)
\\
\displaystyle  A_h^-=\sum_{j=1}^n \e_jA_h^{-j} & \mbox{with}&
A_h^{-j}=\sqrt{\dfrac{qh}{4\mu}}\left(\dfrac{2}{qh}I-\ajminus
T_h^{-j}\right).
\end{array}
\end{eqnarray}

Here we recall that in terms of the identity operator
$I:\f(x)\mapsto \f(x)$ and the forward/backward shifts $T_h^{\pm
j}\f(x)=\f(x\pm h\e_j)$ on the $x_j-$axis, the action $\f(x)\mapsto
L_h \f(x)$ corresponds to
$$
L_h=\frac{1}{2\mu}\sum_{j=1}^n \left(\frac{2}{qh}I-\ajplus
T_h^{+j}-\ajminus T_h^{-j}\right)+q~\Phi_h(x)I.
$$

It is straightforward to verify that
$A_h^{+j}A_h^{-j}+A_h^{-j}A_h^{+j}$
 equals \begin{eqnarray*}-\frac{2}{\mu q h}I+\dfrac{1}{\mu}\ajplus T_h^{+j}+\dfrac{1}{\mu}
\ajminus T_h^{-j}-\dfrac{qh}{4 \mu}
\left(\ajplus^2+\ajminus^2\right)I.
\end{eqnarray*}

Then, for $\displaystyle \Phi_h(x)=\dfrac{h}{8\mu}\sum_{j=1}^n
\left(\ajplus^2+\ajminus^2\right)$ one obtains (see \ref{AhpmLemma})
$$
\frac{1}{2}\left(A_h^+A_h^-+A_h^-A_h^+\right)-q \Phi_h(x)I
=\frac{1}{2\mu}\sum_{j=1}^n \left(\frac{2}{qh}I-\ajplus
T_h^{+j}-\ajminus T_h^{-j}\right),
$$

Thereby, the discrete electric potential $\Phi_h(x)$ is uniquely
determined by the factorization property
$L_h=\dfrac{1}{2}\left(A_h^+A_h^-+A_h^-A_h^+\right)$. On the other
hand, based on the summation formulae (cf. \cite[Subsection
1.5]{MontvayMunster94})
\begin{eqnarray*}
\sum_{x\in h\BZ^n}h^n~ \f(x \pm h\e_j)^\dag \g(x)=\sum_{x\in
h\BZ^n}h^n~ \f(x)^\dag \g(x\mp h\e_j)
\end{eqnarray*}
one easily recognize the following adjoint relations, written in
terms of the shift operators $T_h^{\pm j}$:
\begin{eqnarray}
\label{ajpmAdjoint}
\begin{array}{lll}
\left\langle \ajplus T_h^{+j} \f, \g \right\rangle_h&=&\left\langle
\f, \ajminus T_h^{-j}\g \right\rangle_h \\ \left\langle \ajminus
T_h^{-j} \f, \g \right\rangle_h&=&\left\langle  \f, \ajplus
T_h^{+j}\g \right\rangle_h.
\end{array}
\end{eqnarray}

This yield $A_h^+$ and $A_h^-$ as Hermitian conjugates one of the
other with respect to the Hilbert module $\ell_2(h\BZ^n;\cl_{0,n})$,
as required by the {\bf Duality condition} underlying to the Fock
space $\mathcal{F}_h$ over $h\BZ^n$ (see
\ref{ExactSolvabilityLemma}). Since the {\it vacuum} vector
$\psi_0(x;h)=\phi(x;h)\s$ is $\mbox{Pin}(n)-$valued we can make use
of the method of separation of variables to compute $\phi(x;h)$ from
the set of functional equations
\begin{eqnarray}
\label{VacuumRecursive} \phi(x+h\e_j;h)=\dfrac{2}{q
    h}~\dfrac{1}{a_h(x_j)}\phi(x;h) & \mbox{for each}& j=1,2,\ldots,n.
\end{eqnarray}

Indeed, (\ref{VacuumRecursive}) is equivalent to the set of
equations $A_h^{+j}\phi(x;h)=0$ ($j=1,2,\ldots,n$), and hence, to
$A_h^+\psi_0(x;h)=\left(A_h^+\phi(x;h)\right)\s=0$. Henceforth we
make use of the conjugation property $(\s\f(x))^\dag=\f(x)^\dag
\s^\dag$ to get rid of the Pinor/Spinor element $\s$ on the
quasi-probability formulation (\ref{quasiProbability}) of the {\bf
Energy condition} $\langle \psi_0,\psi_0 \rangle_h=1$. Indeed, for
$\psi_0(x;h)=\phi(x;h)\s$, the quasi-probability law
(\ref{quasiProbability}) carrying a set of {\it independent and
identically distributed} (i.i.d) random variables
$X_1,X_2,\ldots,X_n$ thus becomes
\begin{eqnarray}
\label{quasiProbabilityiid}\Pr\left(\sum_{j=1}^n\e_j X_j=x\right)=h^n\phi(x;h)^2.
\end{eqnarray}

\subsection{Intertwining Properties}

With the aim of obtaining a recovery for the discrete electric and magnetic
 potentials, $\Phi_h(x)$ and
${\bf a}_h(x)$ respectively, from the knowledge of the $k-$bound
states (\ref{PhikXh}) of $L_h$ with membership in the Fock space
$\mathcal{F}_h$, we are going now to establish a general framework
involving a generalization of the quasi-monomiality principle
obtained in author's recent paper \cite{FaustinoMonomiality14}.
 For their construction we shall employ
intertwining properties between $A_h^\pm$ and the set of ladder
Clifford-vector-valued operators
\begin{eqnarray*}
D_h^+&=&\sum_{j=1}^n \e_j\partial_h^{+j} \\
M_h&=&\sum_{j=1}^n \e_j
\left(h\ajminus^2T_h^{-j}-\frac{4}{q^2h}I\right).
\end{eqnarray*}

As usual, $\partial_h^{+j}\f(x)=\dfrac{\f(x+h\e_j)-\f(x)}{h}$
($j=1,2,\ldots,n$) denote the forward finite difference operators on
$h\BZ^n$ (cf. \cite[Subsection 2.1.]{FaustinoMonomiality14}).

First, recall that the {\it vacuum} vector $\psi_0(x;h)=\phi(x;h)\s$
annihilated by $A_h^+$, may be computed from the set of functional
equations (\ref{VacuumRecursive}). More generally, the set of
constraints (\ref{VacuumRecursive}) provide us a scheme to derive an
intertwining property between the degree-lowering type operator
$A_h^+$ and the finite difference Dirac operator $D_h^+$, seemingly
close to the Rodrigues type formula involving the Clifford-Hermite
polynomials/functions (cf.~\cite[Lemma 3.1]{CFK11}). For every
Clifford-vector-valued function $\f(x)$ we thus have the set of
relations
\begin{eqnarray*}
A_h^+\left( \phi(x;h)\f(x) \right)&=& \sum_{j=1}^n
\e_j~\sqrt{\dfrac{qh}{4\mu}} \left(\ajplus
\phi(x+h\e_j;h)\f(x+h\e_j)
-\frac{2}{ q h}\phi(x;h)\f(x)\right) \\
&=& \sqrt{\frac{h}{{\mu q}}}~\sum_{j=1}^n \e_j~
\phi(x;h)\dfrac{\f(x+h\e_j)-\f(x)}{h}
\\
&=& \sqrt{\frac{h}{{\mu q}}}~\phi(x;h)~ D_h^+ \f(x)
\end{eqnarray*}
that in turn yields the operational formula $$\displaystyle
\phi(x;h)^{-1}A_h^+\left(\phi(x;h)\f(x)\right)=\sqrt{\frac{h}{{\mu
q}}} D_h^+ \f(x).$$

In a similar manner one can derive an intertwining property,
involving the operators $A_h^-$ and $M_h$ if we reformulate the set
of functional equations (\ref{VacuumRecursive}) in terms of the
backward shifts $T_h^{-j}\f(x)=\f(x-h\e_j)$. Thereby, the set of
relations
\begin{eqnarray*}
A_h^-\left( \phi(x;h)\f(x) \right) &=&\sum_{j=1}^n
\e_j\sqrt{\dfrac{qh}{4\mu}}
\left(\frac{2}{q h}\phi(x;h)\f(x)-\ajminus \phi(x-h\e_j;h)\f(x-h\e_j)\right) \\
&=&-\sqrt{\dfrac{qh}{4\mu}}~\sum_{j=1}^n \e_j
\phi(x;h)\left(\frac{q h}{2}\ajminus^2 \f(x-h\e_j)-\frac{2}{qh}\f(x)\right)\\
 &=& -\frac{1}{4}\sqrt{\dfrac{q^3h}{\mu}}~\phi(x;h) M_h \f(x),
\end{eqnarray*}
that hold for an arbitrary Clifford-vector-valued function $\f(x)$,
yield as a direct consequence of
\begin{eqnarray*}
\phi(x-h\e_j;h)= \frac{qh}{2}~\ajminus~\phi(x;h) & (j=1,2,\ldots,n).
\end{eqnarray*}

This implies
\begin{eqnarray*}
\displaystyle
\phi(x;h)^{-1}A_h^-(\phi(x;h)\f(x))=-\frac{1}{4}\sqrt{\dfrac{q^3h}{\mu}}~
M_h~\f(x).
\end{eqnarray*}

Furthermore, induction over $k\in \BN_0$ shows that the $k-$bound
states (\ref{PhikXh}) are thus characterized by the operational
formula
\begin{eqnarray}
\label{CliffordMk}
\psi_k(x;h)=\frac{(-1)^k}{4^k}~\frac{q^{\frac{3k}{2}}h^{\frac{k}{2}}}{\mu^{\frac{k}{2}}}~\phi(x;h)~(M_h)^k\s.
\end{eqnarray}

On the other hand, combination of the previously obtained relations
give rive to
\begin{eqnarray*}
\displaystyle \phi(x;h)^{-1} A_h^-A_h^+\left(\phi(x;h)\f(x)\right)
=-\frac{qh}{4\mu}~M_h D_h^+\f(x) \\
\displaystyle \phi(x;h)^{-1} A_h^+A_h^-\left(\phi(x;h)\f(x)\right)
=-\frac{qh}{4\mu}~D_h^+M_h\f(x).
\end{eqnarray*}

This results into the following lemma:
\begin{lemma}\label{quasiMonomialLemma}
Let $\s\in \mbox{Pin}(n)$, $\phi(x;h)$ a scalar-valued function
satisfying (\ref{VacuumRecursive}), $\psi_k(x;h)$ the $k-$bound
states defined {\it viz} equation (\ref{PhikXh}) and
$$ \m_k(x;h)=(M_h)^k\s$$
be quasi-monomials of order $k$ ($k\in\BN_0$). Then we have the
following:
\begin{enumerate}
\item The quasi-monomials
$\m_k(x;h)$ may be determined through the formula
$$\m_k(x;h)={(-1)^k4^k}~
\dfrac{\mu^{\frac{k}{2}}}{q^{\frac{3k}{2}}h^{\frac{k}{2}}}~\dfrac{\psi_k(x;h)}{\phi(x;h)}.$$
\label{quasiMonomialsStatement}
\item The quasi-monomials $\m_k(x;h)$ and the
Fock states $\psi_k(x;h)$ are interrelated by the isospectral
formula
$$\displaystyle
M_hD_h^+\m_{k}(x;h)+D_h^+M_h \m_k(x;h)=
{(-1)^{k+1}4^{k+1}}~\dfrac{\mu^{\frac{k}{2}+1}}{q^{\frac{3k}{2}+1}h^{\frac{k}{2}+1}}~\phi(x;h)^{-1}L_h
\psi_k(x;h).$$\label{IntertwiningStatement}
\end{enumerate}
\end{lemma}

 Regardless the
formal computation of the $\m_k'$s, we would like to stress that the
operator $\left(M_h\right)^{2r}$ $(k=2r)$ is scalar-valued operator
whereas $\left(M_h\right)^{2r+1}$ is vector-valued ($k=2r+1$). To
fill this gap, the computation of the quasi-monomials $\m_{k}(x;h)$
of even and odd orders separately\footnote{This is similarly to what
was done in \cite[Example 3.2]{FaustinoMonomiality14} and in
\cite[Example 3.3]{FaustinoMonomiality14} to compute hypercomplex
versions for the falling factorials and Poisson-Charlier
polynomials, respectively.}. For the even orders ($k=2r$) we use the
multinomial formula
\begin{eqnarray}
\label{PhikXhMultinomial}
\begin{array}{lll}
\m_{2r}(x;h)&=&\left(\left(M_h\right)^{2}\right)^r\s
\\ \ \\
&=& \displaystyle \sum_{|\sigma|=r} (-1)^r \frac{r!}{\sigma!}
\prod_{j=1}^n\left(h\ajminus^2 T_{h}^{-j}
-\frac{4}{q^2h}I\right)^{2\sigma_j}\s
\end{array}
\end{eqnarray}
that results from the operational identity
\begin{eqnarray*} \left(M_h\right)^2&=&-\sum_{j=1}^n
\left(h\ajminus^2T_h^{-j}-\frac{4}{q^2h}I\right)^2,
\end{eqnarray*}
whereas for the odd orders ($k=2r+1$), we take into account the
recursive formula
\begin{eqnarray}
\label{PhikXhRecursive}
  \m_{2r+1}(x;h)=M_h\m_{2r}(x;h).
\end{eqnarray}

Here and elsewhere, for a given multi-index
$\sigma=(\sigma_1,\sigma_2,\ldots,\sigma_n)$, $\displaystyle
|\sigma|=\sum_{j=1}^n\sigma_j$ denotes the multi-index degree and
$\displaystyle \sigma!=\prod_{j=1}^n \sigma_j!$ the multi-index
factorial.

From the construction considered previously it follows from a short
computation that for a given {\it vacuum} vector of the form
$\psi_0(x;h)=\phi(x;h)\s$ ($\s\in \mbox{Pin}(n)$) the discrete
electric and magnetic potentials, $\Phi_h(x)$ and ${\bf a}_h(x)$
respectively, are uniquely determined by the formulae
\begin{eqnarray}\label{Phih}
\displaystyle \Phi_h(x)&=& \frac{h}{8\mu} \sum_{j=1}^n
\frac{4}{q^2h^2}\left(\dfrac{\phi(x;h)^2}{\phi(x+h\e_j;h)^2}+
\dfrac{\phi(x-h\e_j;h)^2}{\phi(x;h)^2}\right)\\
\label{ahj}
 \displaystyle {\bf a}_h(x)
 &=&\sum_{j=1}^n \e_j \frac{2}{qh}~\dfrac{\phi(x;h)}{\phi(x+h\e_j;h)} .
\end{eqnarray}

 This readily
solves part of the question posed in {\bf Problem
\ref{ExactSolvabilityProblem}}. More generally, statement
\ref{quasiMonomialsStatement}. of Lemma \ref{quasiMonomialLemma}
allows us to give a faithful answer to {\bf Problem
\ref{ExactSolvabilityProblem}} as a some sort of inverse problem:

\begin{proposition}\label{AnswerMagnetic}
Let us assume that the $k-$Fock states $\psi_k(x;h)$ of the discrete
electromagnetic Schr\"odinger operator $L_h$ are
$\mbox{Pin}(n)-$valued.

If for a given sequence $\{\m_{k}(x;h)~:~k\in\BN_0 \}$ of
quasi-monomials the statement \ref{IntertwiningStatement}. of Lemma
\ref{quasiMonomialLemma} is fulfilled, then the {\it vacuum} vector
$\psi_0(x;h)=\phi(x;h)\s$ ($\s \in \mbox{Pin}(n)$) may be recovered
from the projection-based formula
$$\phi(x;h)={(-1)^k4^k}~{\dfrac{\mu^{\frac{k}{2}}}{q^{\frac{3k}{2}}h^{\frac{k}{2}}}}~\dfrac{\m_k(x;h)^\dag\psi_k(x;h)}
{\m_k(x;h)^\dag \m_k(x;h)}.$$

Moreover, the discrete electric and magnetic potentials, $\Phi_h(x)$
resp. ${\bf a}_h(x)$, are uniquely determined by inserting the
right-hand side of $\phi(x;h)$ on the formulae (\ref{Phih}) and
(\ref{ahj}), respectively.
\end{proposition}

On the above characterizations, the scalar-valued potential
$\Phi_h(x)$ is determined from the components of the discrete
magnetic potential ${\bf a}_h(x)$ or from the knowledge of the {\it
vacuum} vector. Concerning the construction of the quasi-monomials
we would like to stress that for each $\s \in \mbox{Pin}(n)$
$\m_k(x;h)\s^\dag$ is scalar-valued for $k$ even and vector-valued
for $k$ odd. In particular, for $k=1$
\begin{eqnarray}
\m_1(x+h\e;h)\s^\dag= \sum_{j=1}^n
\e_j\left(h\ajplus^2-\dfrac{4}{q^2h}\right),
&\mbox{with} & \displaystyle \e=\sum_{j=1}^{n}\e_j. 
\end{eqnarray}

A short computation involving the bilinear form $\displaystyle
\mathcal{B}(x,y)=-\frac{1}{2}(xy+yx)$ gives
\begin{eqnarray*}
\mathcal{B}\left(
\frac{1}{h}\m_1(x;h)\s^\dag,\e_j\right)=\ajminus^2-\frac{4}{q^2h^2}&
\mbox{and} &\mathcal{B}\left(
\frac{1}{h}\m_1(x+h\e;h)\s^\dag,\e_j\right)=\ajplus^2-\frac{4}{q^2h^2}.
\end{eqnarray*}

That allows us to formulate the following proposition:
\begin{proposition}\label{AnswerMagneticM1}
Let us assume that $\m_1(x;h)$ is a $\BC\otimes\mbox{Pin}(n)$-valued
quasi-monomial of order $1$.
 Then, the discrete electric and magnetic potentials, $\Phi_h(x)$ and ${\bf a}_h(x)$
respectively, may be recovered from the formulae
\begin{eqnarray*}
\label{MhElectric}
\displaystyle\Phi_h(x)&=&\frac{1}{8\mu}\mathcal{B}\left(\frac{1}{h}
\m_1(x+h\e;h)\s^\dag,\e\right)+\frac{1}{8\mu}\mathcal{B}\left(
\frac{1}{h}\m_1(x;h)\s^\dag,\e\right)+\frac{n}{\mu q^2h^2} \\
\ \\ \label{MhMagnetic} \displaystyle {\bf a}_h(x)&=&\sum_{j=1}^n
\e_j \sqrt{\mathcal{B}\left(\frac{1}{h}
    \m_1(x+h\e;h)\s^\dag,\e_j\right)+\frac{4}{q^2h^2}}.
\end{eqnarray*}

\end{proposition}

Proposition \ref{AnswerMagneticM1} provides an alternative way to
recover the electric and magnetic potentials from the knowledge of
the quasi-monomial of order $k=1$, in interplay with the recursive
formula (\ref{PhikXhRecursive}) for $r=0$.

\section{The Bayesian Probability Insight}\label{BayesianProbabilitySection}

\subsection{Poisson and Hypergeometric Distributions}\label{PoissonHypergeometricSection}

 Our next step is to study the quasi-exact solvability of the
multidimensional discrete electromagnetic Schr\"odinger operator
(\ref{JacobiOperator}) through the connection between the bound
states (\ref{PhikXh}) and the discrete electric and magnetic
potentials. In view of Proposition \ref{AnswerMagnetic} we will
restrict ourselves to the construction of the discrete electric and
magnetic potentials, $\Phi_h(x)$ and ${\bf a}_h(x)$ respectively,
from the knowledge of the ground state $\psi_0(x;h)=\phi(x;h)\s$.

Based on the descriptions (\ref{Phih}) and (\ref{ahj}) obtained in
Section \ref{BoundSection} for $\Phi_h(x)$ and ${\bf a}_h(x)$
respectively, it remains natural to exploit the Fock space
$\mathcal{F}_h$ from the Bayesian probability side
(cf.~\cite{CavesFuchsSchack02,Mouayn14}) by means of the {\it
likelihood} function $x\mapsto h^n \phi(x;h)^2$ encoded by the
quasi-probability law (\ref{quasiProbabilityiid}). Particular
examples arising this construction are\footnote{See also
\cite[Section 4]{Mouayn14}.}:
\begin{enumerate}
\item The multi-variable Poisson-Charlier polynomials, determined from
the multi-variable Poisson distribution with parameter $\lambda>0$
(cf. \cite[p.~335]{FLtVinet93}):
$$h^n\phi(x;h)^2=\left\{
\begin{array}{lll}
\displaystyle \prod_{j=1}^n
\exp(-\lambda)~\dfrac{\lambda^{\frac{x_j}{h}}}{\Gamma\left(\frac{x_j}{h}+1
\right)} &, \mbox{if}~~x \in h\BZ^n_{\geq 0} \\ \ \\
0 &, \mbox{otherwise}
\end{array}\right.$$
\item The multi-variable Meixner polynomials, determined from
the multivariable hypergeometric distribution, defined as (cf.
\cite[pp.~337-338]{FLtVinet93})
$$h^n\phi(x;h)^2=\left\{
\begin{array}{lll}
\displaystyle \prod_{j=1}^n
\dfrac{\Gamma\left(\beta+\frac{x_j}{h}\right)}{\Gamma(\beta)}~
\dfrac{\lambda^{\frac{x_j}{h}}(1-\lambda)^\beta}{\Gamma\left(\frac{x_j}{h}+1
\right)} &, \mbox{if}~~x \in h\BZ^n_{\geq 0} \\ \ \\
0 &, \mbox{otherwise}
\end{array}\right.$$
carrying the parameters $\beta>0$ and $0<\lambda<1$.
\end{enumerate}

For the multi-variable Poisson distribution with parameter
$\lambda=\dfrac{4}{qh^2}$
\begin{eqnarray*}
\displaystyle \Phi_h(x)= \frac{h}{8\mu}\sum_{j=1}^n
\left(\frac{2}{q}~\frac{x_j}{h}+\frac{1}{q}\right)&\mbox{and}& {\bf
a}_h(x)= \sum_{j=1}^n
\e_j~\sqrt{\frac{1}{q}~\frac{x_j}{h}+\frac{1}{q}}
\end{eqnarray*}
are the underlying discrete electric and magnetic potentials,
respectively, defined for $x\in h\BZ^n_{\geq 0}$. Thus, the
Clifford-vector-valued polynomials obtained through the operational
action of the multiplication operator
$$M_h=\sum_{j=1}^n \e_j \frac{1}{q}\left(x_jT_h^{-j}-\dfrac{4}{qh}I
\right)$$ are of Poisson-Charlier type (cf. \cite[Example 3.3
]{FaustinoMonomiality14}). Such families of quasi-monomials are
encoded on the pair $\left( D_h^+,M_h\right)$, by means of
Fischer/Fourier duality (cf.~\cite{FR11,Faustino13}).

For the case where electric charge satisfies the condition
$h>\frac{2}{q}$, the above hypergeometric distribution with
parameters
 $\lambda=\dfrac{4}{q^2h^2}$ and $\beta>0$ endow the discrete electric and magnetic potentials
\begin{eqnarray*}
\Phi_h(x)&=&\left\{
\begin{array}{lll}
\displaystyle \dfrac{h}{8\mu}\sum_{j=1}^n
\left(\dfrac{\frac{1}{q}~\frac{x_j}{h}+\frac{1}{q}}{\frac{1}{q}~\frac{x_j}{h}+\frac{\beta}{q}}+
\dfrac{\frac{1}{q}~\frac{x_j}{h}}{\frac{1}{q}~\frac{x_j}{h}+\frac{\beta-1}{qh}}\right)
&,
\mbox{if}~~x \in h\BZ^n_{\geq 0} \\ \ \\
0 &, \mbox{otherwise}
\end{array}\right.
\\ \ \\
{\bf a}_h(x)&=&\left\{
\begin{array}{lll}
    \displaystyle \sum_{j=1}^n \e_j
    \sqrt{\dfrac{\frac{1}{q}~\frac{x_j}{h}+\frac{1}{q}}{\frac{1}{q}~\frac{x_j}{h}+\frac{\beta}{q}}} &,
    \mbox{if}~~x \in h\BZ^n_{\geq 0} \\ \ \\
    0 &, \mbox{otherwise}
\end{array}\right.
\end{eqnarray*}
that in turn yields $\displaystyle M_h=\sum_{j=1}^n \e_j
h~\left(\dfrac{x_j}{x_j+(\beta-1)
h}T_h^{-j}-\dfrac{4}{q^2h^2}I\right)$
 as multiplication operator, acting on $h\BZ^n_{\geq 0}$.

\subsection{Mittag-Leffler Distributions}\label{MittagLefflerSection}

Let us specialize our results in the case where generalized
Mittag-Leffler functions $E_{\alpha,\beta}$ are involved. The
multivariable {\it likelihood} function that generalizes the
Poisson\footnote{i.e. the {\it likelihood} function determined from
the coefficients of the exponential function
$\exp(\lambda)=E_{1,1}(\lambda)$.} distribution is thus given by
\begin{eqnarray} \label{MittagLefflerDistribution}
h^n\phi(x;h)^2=\left\{
\begin{array}{lll}
\displaystyle \prod_{j=1}^n
E_{\alpha,\beta}\left(\dfrac{4}{q^{2-\alpha}h^{2}}\right)^{-1}
\dfrac{4^{\frac{x_j}{h}}q^{\frac{(2-\alpha)x_j}{h}}h^{-\frac{2
x_j}{h}}}{\Gamma\left(\beta+\alpha\frac{x_j}{h}
\right)} &, \mbox{if}~~x \in h\BZ^n_{\geq 0} \\ \ \\
0 &, \mbox{otherwise}
\end{array}\right.
\end{eqnarray}

As a matter of fact,
$$E_{\alpha,\beta}(\lambda)=\sum_{m=0}^\infty \dfrac{\lambda^m}{\Gamma(\beta+\alpha m)}$$
is well defined for $\mbox{Re}(\alpha)>0$, $\mbox{Re}(\beta)>0$
(cf.~\cite[p.~8]{MathaiSaxenaHaubold09}). A short computation
involving the equations (\ref{Phih})
 and (\ref{ahj}) show that the discrete electric and magnetic fields,
  $\Phi_h(x)$ and ${\bf a}_h(x)$ respectively, are given by the
 general formulae

\begin{eqnarray*}
    \displaystyle \Phi_h(x)&=&\dfrac{h}{8\mu}\sum_{j=1}^{n}
    \frac{1}{q^{\alpha}}\left(\dfrac{\Gamma\left(\alpha+\beta+\alpha \frac{x_j}{h}\right)}
    {\Gamma\left(\beta+\alpha \frac{x_j}{h}\right)}+
    \dfrac{\Gamma\left(\beta+\alpha \frac{x_j}{h}\right)}
    {\Gamma\left(\beta-\alpha+\alpha \frac{x_j}{h}\right)}\right)\\ \displaystyle {\bf
        a}_h(x)&=&\sum_{j=1}^n\e_j~ \sqrt{\frac{1}{q^{\alpha}}~\dfrac{\Gamma\left(\alpha+\beta+\alpha \frac{x_j}{h}\right)}
        {\Gamma\left(\beta+\alpha \frac{x_j}{h}\right)}}.
\end{eqnarray*}
or equivalently,
\begin{eqnarray*}
\displaystyle \Phi_h(x)=\dfrac{h}{8\mu}\sum_{j=1}^{n}
\frac{1}{q^{\alpha}}\left(\left(\beta+\alpha
\frac{x_j}{h}\right)_\alpha+\left(\beta-\alpha+\alpha
\frac{x_j}{h}\right)_\alpha\right)&\mbox{and}& \displaystyle {\bf
a}_h(x)=\sum_{j=1}^n\e_j~
\sqrt{\frac{1}{q^{\alpha}}~\left(\beta+\alpha
\frac{x_j}{h}\right)_\alpha},
\end{eqnarray*}
where $\left(a \right)_\alpha = \dfrac{\Gamma\left(\alpha+a\right)}
{\Gamma\left(a\right)}$ denotes the Pochhammer symbol.

Moreover, the multiplication operator $M_h$ is a polynomial-type
operator of order $\alpha$, given by
\begin{eqnarray}
\label{MittagLefflerMh} M_h=\sum_{j=1}^n\e_j
\left(\frac{h}{q^{\alpha}}~\left(\beta-\alpha+
\alpha\frac{x_j}{h}\right)_\alpha~T_h^{-j}-\frac{4}{q^2h}I\right).
\end{eqnarray}

It is clear from a straightforward application of the generalized
Stirling's formula
\begin{eqnarray}
\label{StirlingF} \Gamma(s+z)\sim \exp(s\log(z))~\Gamma(z)
&\mbox{as}& |z|\rightarrow \infty
\end{eqnarray} that $\Phi_h(x)$ and ${\bf a}_h(x)$ admit,
for $h \rightarrow 0$ and $\alpha\in \BR \setminus \{0\}$, the
asymptotic expansions
\begin{eqnarray*}\displaystyle
\Phi_h(x)\sim \frac{h}{4\mu}\sum_{j=1}^n
\exp\left(\alpha\log\left(\dfrac{\alpha}{q}\dfrac{
x_j}{h}\right)\right) & \mbox{and}& {\bf a}_h(x)\sim \sum_{j=1}^n
\e_j \exp\left({\frac{\alpha}{2}}\log\left(\dfrac{\alpha}{q}\dfrac{
x_j}{h}\right)\right)
\end{eqnarray*}
so that we are under the conditions of eq. (\ref{Hamiltonian}). The
the discrete electromagnetic Schr\"odinger operator is thus
asymptotically equivalent to the Sturm-Liouville operator
$$\f(x) \mapsto -\frac{h^2}{2\mu}\sum_{j=1}^n
 \frac{\partial}{\partial x_j}\left(\exp\left({\frac{\alpha}{2}}\log\left(\dfrac{\alpha}{q}\dfrac{
x_j}{h}\right)\right)\frac{\partial \f}{\partial x_j}(x)\right)
 +V\left(\frac{x}{h}\right)\f(x),$$
with potential $$\displaystyle
V\left(\frac{x}{h}\right)=\frac{1}{2\mu}\sum_{j=1}^n\left[\frac{2}{qh}+\frac{h}{2}\exp\left(\alpha\log\left(\dfrac{\alpha}{q}\dfrac{
x_j}{h}\right)\right)
-\exp\left({\frac{\alpha}{2}}\log\left(\dfrac{\alpha}{q}\dfrac{
x_j}{h}\right)\right)-\exp\left({\frac{\alpha}{2}}\log\left(\dfrac{\alpha}{q}\dfrac{
x_j}{h}-\dfrac{\alpha}{q}\right)\right)\right].$$
%

\subsection{Generalized Wright distributions}
\label{WrightSection}

Widely speaking, one can construct generalizations of the
Mittag-Leffler's distribution (\ref{MittagLefflerDistribution}) by
means of the following Mellin-Barnes integral representation
\begin{eqnarray}
\label{WrightFunctionpq} \displaystyle {~}_p\Psi_t
\left[\begin{array}{ll|} (a_k,\alpha_k)_{1,p} &  \\
(b_l,\beta_l)_{1,t} &
\end{array} ~ \lambda \right]=\dfrac{1}{2\pi i}
\int_{c-i\infty}^{c+i\infty}
\dfrac{\Gamma(s)\prod_{k=1}^p\Gamma(a_k-\alpha_ks)}{\prod_{l=1}^t\Gamma(b_l-\beta_l
s)}(-\lambda)^{-s}~ds.
\end{eqnarray}

Such kind of integral representation formulae correspond to
$H-$function representations of a generalized Wright function, with
parameters $\lambda\in\BC$, $a_k,b_l\in\BC$ and $\alpha_k,\beta_l
\in \BR \setminus \{ 0\}$ ($k=1,2,\ldots,p$; $l=1,2,\ldots,t$) --
see, for instance, \cite[Subsection 1.19]{ErdelyiMagnusEtAll53},
\cite[Subsection 5]{KilbasSaigoTrujillo02} and \cite[Section
1.9]{MathaiSaxenaHaubold09}. Here we notice that in case of the
closed path joining the endpoints $c-i\infty$ and $c+i\infty$
($0<c<1$) contains the simple poles $s=-m$ ($m\in \BN_0$) on the
left, from standard arguments of residue theory, there
holds\footnote{See also \cite[Section 6]{KilbasSaigoTrujillo02}.}
\begin{eqnarray*}
\dfrac{1}{2\pi i}\int_{c-i\infty}^{c+i\infty}
\dfrac{\Gamma(s)\Gamma(1-s)(-\lambda)^{-s}}{\Gamma(\beta-\alpha
s)}~ds &=&\sum_{m=0}^\infty \lim_{s\rightarrow
-m}(s+m)\dfrac{\Gamma(s)\Gamma(1-s)(-\lambda)^{-s}}{\Gamma(\beta-\alpha
s)}
\\
 &=&\sum_{m=0}^\infty
\dfrac{\lambda^{m}}{\Gamma(\beta+\alpha m)},
\end{eqnarray*}
that is $E_{\alpha,\beta}(\lambda)={~}_1\Psi_1
\left[\begin{array}{ll|} (1,1) &  \\
(\beta,\alpha) &
\end{array} ~ \lambda \right]$ (cf. \cite[Example 1.4
]{MathaiSaxenaHaubold09}).

More generally, one can compute generalized multivariable
probability distributions of Wright type, by rewritting
(\ref{WrightFunctionpq}) as a series representation with
coefficients (cf.~\cite[Section 4]{KilbasSaigoTrujillo02} and
\cite[Subsection 1.9.1]{MathaiSaxenaHaubold09})
$$\mu_m=\dfrac{\prod_{k=1}^p\Gamma(a_k+\alpha_km)}{\prod_{l=1}^t\Gamma(b_l+\beta_l
m)}~\dfrac{\lambda^m}{\Gamma(m+1)}.$$

 Assuming that for
$\alpha_k>0$, $\beta_l>0$ ($k=1,2,\ldots,p$;$l=1,2,\ldots,t$) the
intersection between the simple poles $b_l =-m$ ($m\in \BN_0$) of
$\Gamma(s)$ and the simple poles $\frac{a_k+m}{\alpha_k}$
($k=1,\ldots,p;m\in\BN_0$) of $\Gamma(a_k-\alpha_k s)$
($k=1,\ldots,p$) yields an empty set, i.e.
$\frac{a_k+m}{\alpha_k}\neq -m$, the Mellin-Barnes integral
(\ref{WrightFunctionpq}) admits series expansion
\begin{equation}
\label{WrightSeriespq} {~}_p\Psi_t
\left[\begin{array}{ll|} (a_k,\alpha_k)_{1,p} &  \\
(b_l,\beta_l)_{1,t} &
\end{array} ~ \lambda \right]=\sum_{m=0}^\infty
\mu_m
\end{equation}
whenever $\displaystyle \sum_{l=1}^t\beta_l-\sum_{k=1}^p \alpha_k
\geq -1$ (cf. \cite[Theorem 1]{KilbasSaigoTrujillo02}). Moreover:
\begin{enumerate}
\item In case of $\displaystyle
\sum_{l=1}^t\beta_l-\sum_{k=1}^p \alpha_k>-1$, the series expansion
(\ref{WrightSeriespq}) is absolutely convergent for all $\lambda \in
\BC$.
\item In case of $\displaystyle
\sum_{l=1}^t\beta_l-\sum_{k=1}^p \alpha_k=-1$, the series expansion
(\ref{WrightSeriespq}) is absolutely convergent for all values of
$|z|<\rho$ and of $|z|=\rho$, $\mbox{Re}(\mu)>\frac{1}{2}$, with
\begin{eqnarray*}
\displaystyle \rho=\dfrac{\Pi_{l=1}^t
|\beta_l|^{\beta_l}}{\Pi_{k=1}^p |\alpha_k|^{\alpha_k}} & \mbox{and}
& \mu=\sum_{l=1}^t b_l-\sum_{k=1}^p a_k+\frac{p-t}{2}.
\end{eqnarray*}
\end{enumerate}

Therefore, the {\it likelihood} function $x \mapsto h^n\phi(x;h)^2$,
defined {\it viz}
\begin{eqnarray}
\label{WrightDistribution}
\left\{
\begin{array}{lll}
\displaystyle \prod_{j=1}^n {~}_p\Psi_t
\left[\begin{array}{l|l} (a_k,\alpha_k)_{1,p}   \\
(b_l,\beta_l)_{1,t}
\end{array} ~ \lambda \right]^{-1}
\dfrac{\prod_{k=1}^p\Gamma\left(a_k+\alpha_k\frac{x_j}{h}\right)}
{\prod_{l=1}^t\Gamma\left(b_l+
\beta_l\frac{x_j}{h}\right)}~\dfrac{\lambda^{\frac{x_j}{h}}}
{\Gamma\left(\frac{x_j}{h}+1\right)} &,
 \mbox{if}~~x \in h\BZ^n_{\geq 0} \\ \ \\
0 &, \mbox{otherwise}
\end{array}\right.
\end{eqnarray}
corresponds to a Bayesian probability distribution of Wright type.

 Such construction is far beyond the
Mittag-Leffler's distribution (\ref{MittagLefflerDistribution})
since it also encompasses the multi-variable hypergeometric
distribution considered in Subsection
\ref{PoissonHypergeometricSection} (take, for instance, $p=1,t=1$,
$a_1=b_1=\beta$, $\alpha_1=1$ and $\beta_1=0$ on the above formula).
Widely speaking, a wise application of Gauss-Legendre multiplication
formula (cf.~\cite[p.~4]{ErdelyiMagnusEtAll53})
\begin{eqnarray}
\label{GaussLegendre} \prod_{r=0}^{s-1}
\Gamma\left(\frac{r}{s}+z\right)=
(2\pi)^{\frac{s-1}{2}}s^{\frac{1}{2}-s
    z}\Gamma(sz)
\end{eqnarray}
allows us to amalgamize Mittag-Leffler and hypergeometric
distributions as MacRobert's E-functions in disguise
(cf.~\cite[p.~203]{ErdelyiMagnusEtAll53}).

Let us now take, for each $k=1,2,\ldots,\gamma$ ($p=\gamma$) and
$l=1,2,\ldots,\alpha$ ($t=\alpha$), the substitutions
$\alpha_k=\beta_l=1$,
 $\displaystyle a_k=\frac{k-1+\delta}{\gamma}$ and $\displaystyle b_l=\frac{l-1+\beta}{\alpha}$. A straightforward application of (\ref{GaussLegendre}) shows that the coefficients $\mu_m$ of (\ref{WrightSeriespq}) are equal to
 $$ \mu_m=\sqrt{\frac{\gamma}{\alpha}(2\pi)^{\gamma-\alpha}}~
 \dfrac{\alpha^{\beta+\alpha m}\Gamma\left(\delta+\gamma m\right)}
 {\gamma^{\delta+\gamma m}\Gamma\left(\beta+\alpha m\right)}~
 \dfrac{\lambda^m}
 {\Gamma\left(m+1\right)}.$$

Thus, under the condition $\alpha-\gamma> -1$ the {\it likelihood}
function (\ref{WrightDistribution}), carrying the parameter
$\displaystyle
\lambda=\dfrac{\gamma^\gamma}{\alpha^\alpha}\frac{4}{q^{1+\gamma-\alpha}h^{2}}$
simplifies to
\begin{eqnarray}
\label{WrightDistribution11} \left\{
\begin{array}{lll}
\displaystyle \prod_{j=1}^n {~}_1\Psi_1
\left[\begin{array}{l|l} (\delta,\gamma)   \\
(\beta,\alpha)
\end{array} ~\dfrac{\gamma^\gamma}{\alpha^\alpha}\frac{4}{q^{1+\gamma-\alpha}h^{2}} \right]^{-1}
~ \dfrac{ \Gamma\left(\delta+\gamma\frac{x_j}{h}\right)}
{\Gamma\left(\beta+\alpha\frac{x_j}{h}\right)}~
\dfrac{\alpha^{\frac{\alpha x_j}{h}}\gamma^{-\frac{\gamma
x_j}{h}}4^{\frac{x_j}{h}}q^{-\frac{(1+\gamma-\alpha)x_j}{h}}h^{-\frac{2x_j}{h}}}
{\Gamma\left(\frac{x_j}{h}+1\right)} ,&
x \in h\BZ^n_{\geq 0} \\ \ \\
0 &,\mbox{otherwise}
\end{array}\right..
\end{eqnarray}

The above {\it likelihood} function is also well defined\footnote{In
case of $\alpha-\gamma=-1$, the Wright series ${~}_1\Psi_1
\left[\begin{array}{l|l} (\delta,\gamma)   \\
(\beta,\alpha)
\end{array} ~\lambda \right]$
is also absolutely convergent for
$|\lambda|<\dfrac{\alpha^\alpha}{\gamma^\gamma}$ and of
$|\lambda|=\dfrac{\alpha^\alpha}{\gamma^\gamma}$,
$\mbox{Re}(\beta)-\mbox{Re}(\delta)>\frac{1}{2}$.} for $
h^2>\dfrac{\gamma^{2\gamma}}{\alpha^{2\alpha}}~\dfrac{4}{q^{{1+\gamma-\alpha}}}$
and of
$h^2=\dfrac{\gamma^{2\gamma}}{\alpha^{2\alpha}}~\dfrac{4}{q^{{1+\gamma-\alpha}}}$,
$\mbox{Re}(\beta)-\mbox{Re}(\delta)>\frac{1}{2}$ whenever
$\alpha-\gamma=-1$.
In particular:\\
\begin{enumerate}
\item For $
h^2=\dfrac{\gamma^{2\gamma}}{\alpha^{2\alpha}}~\dfrac{4}{q^{{1+\gamma-\alpha}}}$
and
$0<q^{{1+\gamma-\alpha}}<4\dfrac{\gamma^{2\gamma}}{\alpha^{2\alpha}}$,
the above set of formulae are also true under the choice
$\delta=\dfrac{\beta}{2}
=\dfrac{\gamma^{2\gamma}}{\alpha^{2\alpha}}~\dfrac{2}{q^{{1+\gamma-\alpha}}}$.
\item For $\gamma=\delta=1$, the {\it likelihood} function
(\ref{WrightDistribution11}) is the Mittag-Leffler distribution
(\ref{MittagLefflerDistribution}) in disguise. Moreover, if
$\alpha=\mbox{Re}(\alpha)>0$, $\alpha\rightarrow 0^+$ and
$h>\dfrac{2}{q}$, (\ref{MittagLefflerDistribution}) simplifies to
\begin{eqnarray}
\label{MittagLefflerDAsymptotic} h^n\phi(x;h)^2= \left\{
\begin{array}{lll}
\displaystyle \prod_{j=1}^n
\left({1-\frac{4}{q^2h^2}}\right)^{-1}q^{-\frac{2x_j}{h}}h^{-\frac{2x_j}{h}}
&, \mbox{if}~~x \in h\BZ^n_{\geq 0} \\ \ \\
0 &, \mbox{otherwise}
\end{array}\right..
\end{eqnarray}
\item For $\beta=\delta$, the {\it likelihood} function
(\ref{WrightDistribution11}) amalgamates the Poisson distribution
($\alpha=\gamma=1$) as well as the orthogonal measure that gives
rise, up to the constant $\left(1-\frac{4}{q^2h^2}\right)^{-\beta
n}$, to the hypergeometric distribution on $h\BZ^n_{\geq 0}$,
carrying the parameter $\lambda=\frac{4}{q^2h^2}$
($\alpha\rightarrow 0^+$, $\gamma=1$ and $h>\frac{2}{q}$) (see
Subsection \ref{PoissonHypergeometricSection}).
\end{enumerate}

\bigskip

Let us now turn our attention to the construction of quasi-monomials
carrying (\ref{WrightDistribution11}). A direct consequence of
(\ref{PhikXhMultinomial}) shows that the even powers represented
through the operational formula
\begin{eqnarray*}\m_{2r}(x;h)=(-1)^r\left(\sum_{j=1}^n\left(
\frac{h}{q^{1+\alpha-\gamma}}~\frac{x_j}{h}~\frac{\left(\beta-\alpha+
    \alpha\frac{x_j}{h}\right)_\alpha}{\left(
    \delta-\gamma+\gamma\frac{x_j}{h}\right)_\gamma}
    ~T_h^{-j}-\frac{4}{q^2h}I\right)^2\right)^r
    \s&
(~\s \in \mbox{Pin}(n)~)
\end{eqnarray*}
 have the hypergeometric series representation (cf.~\ref{MittagLefflerLemma})
    \begin{eqnarray*}
        \m_{2r}(x;h)=
        (-1)^r
\frac{16^r}{q^{2r(1+\alpha-\gamma)}h^{2r}}\sum_{|\sigma|=r}
        \frac{r!}{\sigma!}
        \prod_{j=1}^n{\bf
        w}_{\sigma_j}(x;h)\s,
    \end{eqnarray*}
    with

${\bf
    w}_{\sigma_j}(x;h)={~}_{2+\alpha}F_\gamma\left(-2\sigma_j,-\frac{x_j}{h},
    \left(\frac{k-1+\beta}{\alpha}-1+\frac{x_j}{h}\right)_{1,\alpha};
    \left(\frac{l-1+\delta}{\gamma}-1+\frac{x_j}{h}\right)_{1,\gamma};
    -\frac{\alpha^\alpha}{\gamma^\gamma}
    \frac{q^{1+\gamma-\alpha}h^2}{4}\right).$

Here and elsewhere ${~}_{2+\alpha}F_\gamma$ denotes the generalized
hypergeometric series expansion (cf.~\cite[Chapter
IV]{ErdelyiMagnusEtAll53}) $$\displaystyle
{~}_{\alpha+1}F_\gamma\left(a,b,
\left(c_k\right)_{1,\alpha};\left(d_l\right)_{1,\gamma};
\lambda\right)= \sum_{p=0}^\infty ~(a)_p~(b)_p~
\dfrac{\prod_{k=1}^\alpha \left(c_k\right)_p}{\prod_{l=1}^\gamma
\left(d_l\right)_p}~\dfrac{\lambda^p}{p!}.$$

By applying (\ref{GaussLegendre}) to the product of Pochhammer
coefficients of $\displaystyle {~}_{\alpha+1}F_\gamma$, there holds
\begin{eqnarray*}
    \prod_{k=0}^{\alpha-1} \left(\frac{k+\beta}{\alpha}-1+\frac{x_j}{h}\right)_p
    &=&\prod_{k=0}^{\alpha-1} \dfrac{\Gamma\left(\frac{\beta+k}{\alpha}-1+\frac{x_j}{h}+p\right)}
    {\Gamma\left(\frac{k+\beta}{\alpha}-1+\frac{x_j}{h}\right)} \\
    &=&\alpha^{-\alpha p}\dfrac{\Gamma\left(\beta-\alpha+\alpha \frac{x_j}{h}+\alpha p\right)}
    {\Gamma\left(\beta-\alpha+\alpha\frac{x_j}{h} \right)},
\end{eqnarray*}
and analogously
\begin{eqnarray*}
    \displaystyle
    \prod_{l=0}^{\gamma-1} \left(\frac{l+\delta}{\gamma}-1+\frac{x_j}{h}\right)_p
    &=& \gamma^{-\gamma p}\dfrac{\Gamma\left(\delta-\gamma+\gamma \frac{x_j}{h}+\gamma p\right)}
    {\Gamma\left(\delta-\gamma+\gamma\frac{x_j}{h}\right)}.
\end{eqnarray*}

Thus, from the identities
$(-2\sigma_j)_p=(-1)^p\dfrac{\Gamma(2\sigma_j+1)}{\Gamma(2\sigma_j+1-p)}$
and $\left(
-\frac{x_j}{h}\right)=(-1)^p\dfrac{\Gamma\left(\frac{x_j}{h}+1\right)}{\Gamma\left(\frac{x_j}{h}+1\right)}$
 it results the formal series representation
\begin{eqnarray*}
{~}_{2+\alpha}F_\gamma\left(-2\sigma_j,-\frac{x_j}{h},
    \left(\frac{k-1+\beta}{\alpha}-1+\frac{x_j}{h}\right)_{1,\alpha};
    \left(\frac{l-1+\delta}{\gamma}-1+\frac{x_j}{h}\right)_{1,\gamma};
    -\frac{\alpha^\alpha}{\gamma^\gamma}
    \frac{q^{1+\gamma-\alpha}h^2}{4}\right)=\\=\frac{\Gamma\left(\delta-\gamma+\gamma\frac{x_j}{h}
\right)\Gamma\left(2\sigma_j+1\right)\Gamma\left(\frac{x_j}{h}+1\right)}{\Gamma\left(\beta-\alpha+\alpha\frac{x_j}{h}
\right)}\times \\
\times {~}_1\Psi_3 \left[\begin{array}{llll|l}
\left(\beta-\alpha+\alpha \frac{x_j}{h},\alpha\right) & & &
   \\
\left(\delta-\gamma+\gamma \frac{x_j}{h},\gamma\right)&
&\left(2\sigma_j+1,-1\right)&\left(\frac{x_j}{h}+1,-1\right) &
\end{array} -\frac{q^{1+\gamma-\alpha}h^2}{4} ~\right].
\end{eqnarray*}

Here we would like to stress that the right-hand side of the above
expansion shall be understood as a $2\sigma_j-$term truncation of
the generalized Wright series expansion (\ref{WrightSeriespq}).

 Thus, for $r>0$ the even quasi-monomials
$\m_{2r}(x;h)$ are given by
\begin{eqnarray*}\m_{2r}(x;h)=(-1)^r\frac{16^r}{q^{2r(1+\alpha-\gamma)}h^{2r}}\sum_{|\sigma|=r}
\frac{r!}{\sigma!} \prod_{j=1}^n
\frac{\Gamma\left(\delta-\gamma+\gamma\frac{x_j}{h}
\right)\Gamma\left(2\sigma_j+1\right)}{\Gamma\left(\beta-\alpha+\alpha\frac{x_j}{h}
\right)} \times
\\
\times {~}_1\Psi_3 \left[\begin{array}{llll|l}
\left(\beta-\alpha+\alpha \frac{x_j}{h},\alpha\right) & & &
   \\
\left(\delta-\gamma+\gamma \frac{x_j}{h},\gamma\right)&
&\left(2\sigma_j+1,-1\right)&\left(\frac{x_j}{h}+1,-1\right) &
\end{array} -\frac{q^{1+\gamma-\alpha}h^2}{4}~ \right]\s,~~\mbox{with}~~\s\in\mbox{Pin}(n).
\end{eqnarray*}

This surprisingly subtle characterization may be seen as an
hypercomplex extension of Gauss's hypergeometric representation (cf.
\cite[Subsection
2.1.3.]{ErdelyiMagnusEtAll53}~\&~\cite[p.~24]{MathaiSaxenaHaubold09})
when $\alpha\rightarrow 0^+$ and $\gamma=1$. In the case of
$\gamma\rightarrow 0^+$, the above representation becomes a Kummer's
type representation (cf.~\cite[p.~24]{MathaiSaxenaHaubold09}) that
amalgamizes the hypercomplex extension of the Poisson-Charlier
polynomials ($\alpha =\delta=1$) of even order
 (cf. \cite[Example 3.3]{FaustinoMonomiality14}) as well as
  an hypercomplex counterpart for the Meixner polynomials,
  up to the constant $\left(1-\frac{4}{q^2h^2}\right)^{-\beta n}$
  ($\alpha=1$ \& $\delta=\beta$).

\section{Further remarks on quasi-probabilities}\label{RemarksQuasiP}

Although the examples treated throughout Section
\ref{MittagLefflerSection} and Section \ref{WrightSection} involve
Bayesian probabilities in the classical sense, as the ones
considered by the papers \cite{CavesFuchsSchack02,Mouayn14}, the
framework developed provides us a goal-oriented guide to extend to
Bayesian quasi-probabilities with {\it imaginary} bias such as the
one obtained via Bender-Hook-Meisinger-Wang's approach
(cf.~\cite[Section 3]{BenderHookMeiWang10}).

In this section we illustrate the case of complex-valued {\it
likelihood} functions that give rise to $\BC \otimes
\mbox{Pin}(n)-$valued {\it vacuum} vectors $\psi_0(x;h)=\phi(x;h)\s$
towards the regularization of Mittag-Leffler distribution
(\ref{MittagLefflerDistribution}). To illustrate that let us define
for each $\varepsilon>0$ the family of complex-valued {\it
likelihood} functions $x\mapsto h^n\phi_\varepsilon(x;h)^2$ on
$h\BZ^n$, carrying the parameter $\lambda=
\left(\dfrac{4}{q^{2-\alpha}h^{2}}\right)^{1-\varepsilon}{e^{\frac{i\pi\varepsilon}{2}}}$,
as a quasi-probability distribution
\begin{eqnarray}
\label{MittagLefflerDistributionEps} \left\{\begin{array}{lll}
\displaystyle \prod_{j=1}^n
{~}_3\Theta_3\left(({q^{2}h^{2+\alpha}})^{\varepsilon-1}{e^{\frac{i\pi\varepsilon}{2}}}\right)^{-1}
\dfrac{\sin\left(\frac{\pi \varepsilon x_j}{2
h}\right)}{\alpha^\alpha\sin\left(\frac{\pi \varepsilon
x_j}{2\alpha^\alpha
h}\right)}~\dfrac{4^{\frac{(1-\varepsilon)x_j}{h}}q^{\frac{-(2-\alpha)(1-\varepsilon)x_j}{h}}
h^{\frac{-2(1-\varepsilon)x_j}{h}}} { {\Gamma\left(\beta
e^{i\pi\varepsilon}+\alpha\frac{(1-\varepsilon)x_j}{h}
\right)}}~e^{\frac{i\pi \varepsilon x_j}{2h}} ~
,& x \in h\BZ^n_{\geq 0}\\ \ \\
\displaystyle \prod_{j=1}^n
{~}_3\Theta_3\left(({q^{2}h^{2+\alpha}})^{\varepsilon-1}{e^{\frac{i\pi\varepsilon}{2}}}\right)^{-1}
\dfrac{\sin\left(\frac{\pi \varepsilon x_j}{2
h}\right)}{\alpha^\alpha\sin\left(\frac{\pi \varepsilon
x_j}{2\alpha^\alpha
h}\right)}~\dfrac{\varepsilon^{\frac{x_j}{h}}4^{\frac{-(1-\varepsilon)x_j}{h}}q^{\frac{(2-\alpha)(1-\varepsilon)x_j}{h}}
h^{\frac{2(1-\varepsilon)x_j}{h}}}{ {\Gamma\left(\beta
e^{i\pi\varepsilon}+\alpha\frac{(1-\varepsilon)x_j}{h}
\right)}}~e^{\frac{i\pi \varepsilon x_j}{2h}} ,&\mbox{otherwise}
\end{array}\right.
\end{eqnarray}
endowed by the Laurent series expansion
\begin{eqnarray*}
{~}_3\Theta_3\left(\lambda\right) &=&{~}_3\Psi_3
\left[\begin{array}{lll|l} (1,1) &
\left(1,-\frac{\varepsilon}{2\alpha^\alpha}\right) &
 \left(1,\frac{\varepsilon}{2\alpha^\alpha}\right)  \\
(\beta e^{i\pi\varepsilon},\alpha-\varepsilon\alpha) &
 (1,-\frac{\varepsilon}{2})& (1,\frac{\varepsilon}{2})
\end{array} ~\lambda
\right] +\\
&+&{~}_3\Psi_3 \left[\begin{array}{lll|l} (1,1) &
\left(1,-\frac{\varepsilon}{2\alpha^\alpha}\right) &
 \left(1,\frac{\varepsilon}{2\alpha^\alpha}\right)  \\
(\beta e^{i\pi\varepsilon},\alpha-\alpha\varepsilon) &
 (1,-\frac{\varepsilon}{2})& (1,\frac{\varepsilon}{2})
\end{array} ~\frac{\varepsilon}{\lambda}
\right]-\frac{1}{\Gamma(\beta e^{i\pi\varepsilon})}.
\end{eqnarray*}

 On the above construction we make use of the formulae
$\Gamma(1+z)=z\Gamma(z)$ and
$\Gamma\left(z\right)\Gamma\left(1-z\right)=\dfrac{\pi}{\sin(\pi
z)}$ (cf.~\cite[p.~4]{ErdelyiMagnusEtAll53}). Here we recall that
the {\it likelihood} function (\ref{MittagLefflerDistributionEps})
is well defined:
\begin{enumerate}
\item For every
$\lambda=
\left(\dfrac{4}{q^{2-\alpha}h^{2}}\right)^{1-\varepsilon}{e^{\frac{i\pi\varepsilon}{2}}}\in\BC$
whenever\footnote{The case $0<\varepsilon<1$ follows from the fact
that in case of $(1-\varepsilon)\alpha>0$ the Laurent series
${~}_3\Theta_3(\lambda)$ is absolutely convergent in $\BC$.}
$0<\varepsilon<1$.
\item For\footnote{We can rid the condition $\mbox{Re}(-\beta)>\frac{3}{2}$
that yields from application of \cite[Theorem
1]{KilbasSaigoTrujillo02} because $\Gamma\left(\beta e^{i\pi
\varepsilon}+\alpha\frac{(1-\varepsilon)x_j}{h} \right)$ equals to
the constant $\Gamma(-\beta)$ in case of $\varepsilon=1$.} $\beta
\not \in \BN_0$, in case of $\varepsilon\rightarrow 1^-$.
\end{enumerate}

From the limit property $\displaystyle \lim_{\varepsilon \rightarrow
0^+}\dfrac{\sin\left(\frac{\pi \varepsilon x_j}{2
h}\right)}{\alpha^\alpha \sin\left(\frac{\pi \varepsilon
x_j}{2\alpha^\alpha h}\right)}=1$ there holds
\begin{eqnarray*}
\displaystyle \lim_{\varepsilon \rightarrow 0^+}{~}_3\Psi_3
\left[\begin{array}{lll|l} (1,1) &
\left(1,-\frac{\varepsilon}{2\alpha^\alpha}\right) &
 \left(1,\frac{\varepsilon}{2\alpha^\alpha}\right)  \\
(\beta e^{i\pi\varepsilon},\alpha-\varepsilon\alpha) &
 (1,-\frac{\varepsilon}{2})& (1,\frac{\varepsilon}{2})
\end{array} ~\left(\dfrac{4}{q^{2-\alpha}h^{2}}\right)^{1-\varepsilon}{e^{\frac{i\pi\varepsilon}{2}}}
\right]&=&{~}_1\Psi_1
\left[\begin{array}{ll|} (1,1) &  \\
(\beta,\alpha) &
\end{array} ~ \dfrac{1}{q^{2-\alpha}h^{2}} \right]
\\
 \displaystyle
\lim_{\varepsilon \rightarrow 0^+} {~}_3\Psi_3
\left[\begin{array}{lll|l} (1,1) &
\left(1,-\frac{\varepsilon}{2\alpha^\alpha}\right) &
 \left(1,\frac{\varepsilon}{2\alpha^\alpha}\right)  \\
(\beta e^{i\pi\varepsilon},\alpha-\varepsilon\alpha) &
 (1,-\frac{\varepsilon}{2})& (1,\frac{\varepsilon}{2})
\end{array} ~{\varepsilon}\left(\dfrac{4}{q^{2-\alpha}h^{2}}\right)^{\varepsilon-1}{e^{-\frac{i\pi\varepsilon}{2}}} \right]&=&\frac{1}{\Gamma(\beta)}.
\end{eqnarray*}

Thus, for $0<\varepsilon<1$ (\ref{MittagLefflerDistributionEps}) is
a complex-valued regularization of the Mittag-Leffler distribution
(\ref{MittagLefflerDistribution}), since
$$\displaystyle \lim_{\varepsilon \rightarrow 0^+}
{~}_3\Theta_3\left(\left(\dfrac{4}{q^{2-\alpha}h^{2}}\right)^{1-\varepsilon}{e^{\frac{i\pi\varepsilon}{2}}}
\right)=
{~}_1\Psi_1
\left[\begin{array}{ll|} (1,1) &  \\
(\beta,\alpha) &
\end{array} ~ \dfrac{1}{q^{2-\alpha}h^{2}}
\right].
$$

In case of $\varepsilon\rightarrow 1^-$, it follows that
\begin{eqnarray*}
\label{MittagLefflerDistributionEps1}
\begin{array}{lll}
\displaystyle \lim_{\varepsilon \rightarrow 1^-}
h^n\phi_\varepsilon(x;h)^2&=& \displaystyle \prod_{j=1}^n
{~}_3\Theta_2\left(\lambda\right)^{-1}~ \dfrac{\sin\left(\frac{\pi
x_j}{2 h}\right)}{\alpha^\alpha \sin\left(\frac{\pi
x_j}{2\alpha^\alpha h}\right)}~{e^{\frac{i\pi x_j}{2h}}}
\end{array} (x\in h\BZ^n),
\end{eqnarray*}
with
 \begin{eqnarray*}
{~}_3\Theta_2(\lambda)&=&{~}_3\Psi_2\left[\begin{array}{lll|l} (1,1)
& \left(1,-\frac{1}{2\alpha^\alpha}\right) &
 \left(1,\frac{1}{2\alpha^\alpha}\right)  \\
 (1,-\frac{1}{2})& (1,\frac{1}{2}) &
\end{array} ~{i}
\right]+\\
&+&{~}_3\Psi_2\left[\begin{array}{lll|l} (1,1) &
\left(1,-\frac{1}{2\alpha^\alpha}\right) &
 \left(1,\frac{1}{2\alpha^\alpha}\right)  \\
 (1,-\frac{1}{2})& (1,\frac{1}{2}) &
\end{array} ~{-i}
\right]-1.
\end{eqnarray*}

This quasi-probability like distribution is no longer a
regularization for the Mittag-Leffler distribution
(\ref{MittagLefflerDistribution}). On the other hand, it exhibits a
{\it space-time symmetry}\footnote{A $\mathcal{PT}-$symmetry,
accordingly to nomenclature adopted on the papers
\cite{Bender05,BenderEtAll06,BenderHook08,BenderHookMeiWang10}.} due
to the invariance property
$$\lim_{\varepsilon \rightarrow 1^-}
h^n\overline{\phi_\varepsilon(-x;h)^2}=\lim_{\varepsilon \rightarrow 1^-}
h^n\phi_\varepsilon(x;h)^2.$$

 Interesting enough is that the resulting discrete magnetic
 potential
\begin{eqnarray*}
{\bf a}_h(-ix)&=&\sum_{j=1}^n~-i\e_j \dfrac{\sinh\left(\frac{\pi
x_j}{2\alpha^\alpha
h}+\frac{\pi}{2\alpha^{\alpha}}\right)}{qh\sinh\left(\frac{\pi
x_j}{2\alpha^\alpha h}\right)}~\tanh\left(\frac{\pi x_j}{2h}\right)
\end{eqnarray*}
obtained from the transformation $x\mapsto -ix$ on the formula
(\ref{ahj}) is closely related with the hyperbolic potentials of
Macdonald-Ruijnaars type
(cf.~\cite{VanDiejen05,VanDiejenEmsiz15AnnHenriP}).


\section{Conclusions}\label{ConclusionsSection}
Emphasizing how the use of quasi-probabilities may be useful in the
construction of Fock spaces over lattices, we have developed a
framework on which the $k-$Fock states $\psi_k(x;h)$ of $L_h$, and
moreover, the quasi-monomials $\m_k(x;h)$ can be determined from a
general {\it vacuum} vector of the form $\psi_0(x;h)=\phi(x;h)\s$
($\s \in \mbox{Pin}(n)$), encoded by the quasi-probability law
$\displaystyle \mbox{Pr}\left(\sum_{j=1}^n\e_j
X_j=x\right)=h^n\phi(x;h)^2$. The main novelty here against
\cite{FR11,Faustino13,FaustinoMonomiality14} stems into the
description of families of special functions of hypercomplex
variable through the Bayesian quasi-probability formulation rather
than seeking through the set of underlying symmetries.

We make use of Mellin-Barnes integrals to get in touch with Dirac's
framework on quasi-probabilities \cite{Dirac42}. In the shed of the
$H-$Fox framework, it is not surprising that applications in
statistics may be considered in the context of the presented
approach (cf. \cite[Chapter 4]{MathaiSaxenaHaubold09}). On the other
hand, since the Lagrangian operators from relativistic wave
mechanics encompass conserved current densities that may be
interpreted as quasi-probabilities (cf.~\cite[pp.~5-8]{Dirac42}), we
expect that the Bayesian quasi-probability formalism developed
throughout this paper may be useful to investigate questions in
lattice quantum mechanics towards gauge fields, fermion fields and
Quantum Cromodynamics (cf. \cite[Chapter 3, Chapter 4 \& Chapter
5]{MontvayMunster94}), beyond the applications already considered in
\cite{ChakJeugt10,StoilovaJeugt11,MikiTsuVinetZhed12,Mouayn14}. We
also believe that this approach may be useful to establish a deep
and thorough analysis of quantum field models that exhibit axial
anomalies such as the ones considered in \cite{BenderEtAll06}.

The examples involving $H-$Fox functions -- in concrete, the generalized
Mittag-Leffler $E_{\alpha,\beta}(\lambda)$ and Wright
functions ${~}_p\Psi_t
\left[\begin{array}{ll|} (a_k,\alpha_k)_{1,p} &  \\
(b_l,\beta_l)_{1,t} &
\end{array} ~ \lambda \right]$ -- displays also a tangible
interplay between Mellin-Barnes type integrals and fractional
calculus (cf. \cite[Chapter 3]{MathaiSaxenaHaubold09}). Such
connection seems to have been somehow overlooked by several authors
when they are dealing with families of orthogonal polynomials beyond
the known ones within the Askey-Wilson scheme (cf.
\cite{OdakeSasaki09b,VanDiejenEmsiz15AnnHenriP,VanDiejenEmsiz15AdvMath}).

Due to the lack of applications on the literature concerning the
interplay between Bayesian quasi-probabilities with {\it imaginary}
bias and $\mathcal{PT}$-symmetric quantum mechanics (cf.
\cite{Bender05,BenderEtAll06,BenderHook08,BenderHookMeiWang10}) we
believe that this topic deserves a closer inspection, beyond the
simplest examples considered in \cite[Section
4]{BenderHookMeiWang10} and in Section \ref{RemarksQuasiP}. Further
applications of this approach to crystallographic root systems
towards Macdonald-Ruijnaars (pseudo) Laplacians (cf.
\cite{VanDiejen05}) will also be investigated in depth in a future
research.

%

\appendix

\section{Technical Results used in Section \ref{BoundSection}}




\begin{lemma}\label{AhpmLemma}
    For the pair of Clifford-vector-valued ladder operators
    $(A_h^+,A_h^-)$ defined as
\begin{eqnarray}
\label{Ahpm} \displaystyle A_h^{+}=\sum_{j=1}^n \e_j A_h^{+j} &
\mbox{and} & A_h^{-}=\sum_{j=1}^n \e_j A_h^{-j},
\end{eqnarray}
the anti-commutator
    $A_h^{-}A_h^{+}+A_h^{+}A_h^{-}$ is scalar-valued whenever
    $[A_h^{-k},A_h^{+j}]=0$ for $j \neq k$. Moreover, we have
    \begin{eqnarray*}
        A_h^{-}A_h^{+}+A_h^{+}A_h^{-}
        &=&-\sum_{j=1}^n \left(A_h^{-j}A_h^{+j}+A_h^{+j}A_h^{-j}\right).
    \end{eqnarray*}
\end{lemma}

\proof Starting from the definition, we obtain from
(\ref{CliffordGenerators})
\begin{eqnarray*}
A_h^{-}A_h^{+}+A_h^{+}A_h^{-}
&=&\sum_{j,k=1}^n \left(\e_j \e_k A_h^{-j}A_h^{+k}+ \e_k \e_j A_h^{+k}A_h^{-j}\right) \\
&=&\sum_{j,k=1}^n \left(-2\delta_{jk} A_h^{-j}A_h^{+k}+ \e_k \e_j
[A_h^{+k},A_h^{-j}]\right).
\end{eqnarray*}

We see therefore that the bivector summands $\e_k \e_j
[A_h^{+k},A_h^{-j}]$ of $A_h^{-}A_h^{+}+A_h^{+}A_h^{-}$ vanish only
in case of $[A_h^{+k},A_h^{-j}]=0$ hold for every
$j,k=1,2,\ldots,n$, with $j \neq k$. Thus, we have
\begin{eqnarray*}
A_h^{-}A_h^{+}+A_h^{+}A_h^{-} &=&-2\sum_{j=1}^n
A_h^{-j}A_h^{+j}-\sum_{j=1}^n [A_h^{+j},A_h^{-j}].
\end{eqnarray*}

Finally, from the expression
$\left[A_h^{+j},A_h^{-j}\right]=A_h^{+j}A_h^{-j}-A_h^{-j}A_h^{+j}$
we can see that $-2A_h^{-j}A_h^{+j}-[A_h^{+j},A_h^{-j}]$ equals to
$-A_h^{-j}A_h^{+j}-A_h^{+j}A_h^{-j}$, and hence, the above relation
may also be rewritten as
$$
A_h^{-}A_h^{+}+A_h^{+}A_h^{-}=-\sum_{j=1}^n
\left(A_h^{-j}A_h^{+j}+A_h^{+j}A_h^{-j}\right).
$$
\qed



\begin{lemma}\label{ExactSolvabilityLemma}
    For every $\f$ and $\g$ with membership in
    $\ell_2(h\BZ^n;\cl_{0,n})$ there holds
    \begin{eqnarray*}
        \langle A_h^+ \f,\g \rangle_h=\langle \f,A_h^-\g \rangle_h &
        \mbox{and}& \langle A_h^- \f,\g \rangle_h=\langle \f,A_h^+\g
        \rangle_h.
    \end{eqnarray*}

    Moreover
    $$\langle L_h\f,\g \rangle_h
    =\langle \f, L_h\g\rangle_h=\dfrac{1}{2}\langle A_h^+\f,A_h^+\g \rangle_h+\dfrac{1}{2}\langle A_h^-\f,A_h^-\g \rangle_h.$$
\end{lemma}

\proof
Recall that from the $\dag-$ conjugation
properties
 $\displaystyle \left(\e_j A_h^{\pm j} \f(x)\right)^\dag=-(A_h^{\pm j} \f(x))^\dag
 \e_j,$ that follow from (\ref{dagConjugation}), we obtain for each $j=1,2,\ldots,n$,
 the conjugation formula
 $$\displaystyle \left(A_h^{\pm}\f(x)\right)^\dag=-\sum_{j=1}^n \left(A_h^{\pm j} \f(x)\right)^\dag
 \e_j.$$

On the other hand, from (\ref{ajpmAdjoint}) we find that the ladder
operators $A_h^{\pm j}$ defined {\it viz} (\ref{Ahpm})
 satisfy $
\langle A_h^{+j} \f, \g \rangle_h=-\langle  \f, A_h^{-j}\g \rangle_h
$ and $ \langle A_h^{-j} \f, \g \rangle_h=-\langle  \f, A_h^{+j}\g
\rangle_h$.

Combination of the above properties results, for each
$j=1,2,\ldots,n$, into the sequence of relations:
\begin{eqnarray*}
\langle \e_j A_h^{\pm j}\f(x),\g(x) \rangle_h&=& -\langle A_h^{\pm
j}\f(x),\e_j\g(x) \rangle_h \\
&=&
 \langle \f(x),\e_jA_h^{\mp j}\g(x) \rangle_h.
\end{eqnarray*}

Hence, the Hermitian conjugation properties
\begin{eqnarray*}
\label{AhpmDuality} \langle A_h^+ \f, \g \rangle_h=\langle  \f,
A_h^-\g \rangle_h & \mbox{and} & \langle A_h^- \f, \g
\rangle_h=\langle \f, A_h^+\g \rangle_h.
\end{eqnarray*}
in $\ell_2(h\BZ^n;\cl_{0,n})$, and moreover, the set of identities $$\langle L_h\f,\g \rangle_h
=\langle \f, L_h\g\rangle_h=\dfrac{1}{2}\langle A_h^+\f,A_h^+\g \rangle_h+\dfrac{1}{2}\langle A_h^-\f,A_h^-\g \rangle_h$$
 follow straightforwardly from the factorization formula $\displaystyle L_h=\frac{1}{2}(A_h^+ A_h^-+A_h^-A_h^+)$.
%
%
 \qed

\section{Technical Results used in Section \ref{BayesianProbabilitySection}}

\begin{lemma}\label{MittagLefflerLemma}
    In case of $x\mapsto h^n \phi(x;h)^2$ corresponds to the
    {\it likelihood} function (\ref{WrightDistribution11}),
    we thus have
    \begin{eqnarray*}
\m_{2r}(x;h)= (-1)^r
\frac{16^r}{q^{2r(1+\alpha-\gamma)}h^{2r}}~\sum_{|\sigma|=r}\frac{r!}{\sigma!}\prod_{j=1}^n{\bf
w}_{\sigma_j}(x;h)\s
    \end{eqnarray*}
    with
${\bf
    w}_{\sigma_j}(x;h)={~}_{2+\alpha}F_\gamma\left(-2\sigma_j,-\frac{x_j}{h},
    \left(\frac{k-1+\beta}{\alpha}-1+\frac{x_j}{h}\right)_{1,\alpha};
    \left(\frac{l-1+\delta}{\gamma}-1+\frac{x_j}{h}\right)_{1,\gamma};
    -\frac{\alpha^\alpha}{\gamma^\gamma}
    \frac{q^{1+\gamma-\alpha}h^2}{4}\right).$
    \end{lemma}

\proof A direct computation involving the binomial identity shows
that
\begin{eqnarray*}
    \left(\frac{h}{q^{1+\alpha-\gamma}}~\dfrac{x_j}{h}\dfrac{\left(
   \beta-\alpha+ \alpha\frac{x_j}{h}\right)_\alpha}{\left(
    \delta-\gamma+\gamma\frac{x_j}{h}\right)_\gamma}~T_h^{-j}-\frac{4}{q^2h}I\right)^{2\sigma_j}
    &=&\left(\frac{h}{q^{1+\alpha-\gamma}}\right)^{2\sigma_j}~\sum_{p=0}^{2\sigma_j} \left(\begin{array}{ccc}
        2\sigma_j \\p
    \end{array}\right) \left(-\frac{4}{q^{1+\gamma-\alpha}h^2}\right)^{2\sigma_j-p}
    \times \\
&\times& \left(\dfrac{x_j}{h}~\dfrac{\left(
   \beta-\alpha+ \alpha\frac{x_j}{h}\right)_\alpha}{\left(
    \delta-\gamma+\gamma\frac{x_j}{h}\right)_\gamma}~T_h^{-j}
    \right)^p \\
    &=& \left(\frac{4}{q^{2(1+\alpha-\gamma)}h}\right)^{2\sigma_j}~\sum_{p=0}^{2\sigma_j} \left(\begin{array}{ccc}
        2\sigma_j \\p
    \end{array}\right) \left(-\frac{q^{1+\gamma-\alpha}h^2}{4}\right)^{p} \times
    \\
& \times & (-1)^p\left(-\frac{x_j}{h}\right)_p~
\dfrac{\prod_{k=0}^{p-1}\left(\alpha\frac{x_j}{h}+\beta-(k+1)~\alpha\right)_\alpha}
    {\prod_{l=0}^{p-1}\left(\gamma\frac{x_j}{h}+\delta-(k+1)~\gamma\right)_\alpha}.
\end{eqnarray*}

Here we recall $\left(\begin{array}{ccc}
2\sigma_j \\p
\end{array}\right)=(-1)^p\dfrac{(-2\sigma_j)_p}{p!}$
for $p\leq 2\sigma_j$ and $\left(\begin{array}{ccc}
2\sigma_j \\p
\end{array}\right)=0$ for $p>2\sigma_j$. On the other hand, the product $\displaystyle \prod_{k=1}^p
\left(\beta+\alpha\frac{x_j}{h}-k~\alpha\right)_\alpha$ may be rewritten as
\begin{eqnarray*}
    \prod_{k=0}^{p-1}\left(\beta+\alpha \frac{x_j}{h}-(k+1)\alpha \right)_\alpha&=&
    \prod_{k=0}^{p-1}\prod_{s=0}^{\alpha-1}\left(
    \alpha \left(\frac{\beta}{\alpha}-k-1+\frac{x_j}{h}\right) +s\right) \\
    &=&\alpha^{\alpha p}\prod_{s=0}^{\alpha-1}
    \left(\frac{s+\beta}{\alpha}-1+\frac{x_j}{h}\right)_p,
\end{eqnarray*}
and analogously,
\begin{eqnarray*}
    \prod_{l=0}^{p-1}\left(\delta+\gamma \frac{x_j}{h}-(l+1)\gamma \right)_\gamma&=&
\gamma^{\gamma p}\prod_{s=0}^{\gamma-1}
\left(\frac{s+\delta}{\gamma}-1+\frac{x_j}{h}\right)_p.
\end{eqnarray*}

This implies \begin{eqnarray*}
    \left(\frac{h}{q^{1+\alpha-\gamma}}~\dfrac{x_j}{h}\dfrac{\left(
   \beta-\alpha+ \alpha\frac{x_j}{h}\right)_\alpha}{h^\gamma\left(
    \delta-\gamma+\gamma\frac{x_j}{h}\right)_\gamma}~T_h^{-j}-\frac{4}{q^2h}I\right)^{2\sigma_j}
    &=&
    \left(\frac{4}{q^{2(1+\alpha-\gamma)}h}\right)^{2\sigma_j}~\sum_{p=0}^{\infty}
    \frac{1}{p!}~\left(-\frac{\alpha^\alpha}{\gamma^\gamma}
    \frac{q^{1+\gamma-\alpha}h^2}{4}\right)^{p}\times \\
  &\times& (-2\sigma_j)_p\left(-\frac{x_j}{h}\right)_p~
  \dfrac{\prod_{k=0}^{\alpha-1}
    \left(\frac{k+\beta}{\alpha}-1+\frac{x_j}{h}\right)_p}
    {\prod_{l=0}^{\gamma-1} \left(\frac{l+\delta}{\gamma}-1+\frac{x_j}{h}\right)_p}
    \\
    &=&\left(\frac{4}{q^{2(1+\alpha-\gamma)}h}\right)^{2\sigma_j}{\bf
    w}_{\sigma_j}(x;h),
\end{eqnarray*}
with ${\bf
    w}_{\sigma_j}(x;h)={~}_{2+\alpha}F_\gamma\left(-2\sigma_j,-\frac{x_j}{h},
    \left(\frac{k-1+\beta}{\alpha}-1+\frac{x_j}{h}\right)_{1,\alpha};
    \left(\frac{l-1+\delta}{\gamma}-1+\frac{x_j}{h}\right)_{1,\gamma};
    -\frac{\alpha^\alpha}{\gamma^\gamma}
    \frac{q^{1+\gamma-\alpha}h^2}{4}\right).$

By inserting the above relation on the right-hand side of
(\ref{PhikXhMultinomial}), we obtain for $|\sigma|=r$ the desired
result.
\qed

\end{document}